\documentclass[twocolumns]{aa} %

\usepackage{graphicx}
\usepackage{epstopdf}
\usepackage{deluxetable}
\usepackage{times,epsfig} 
\usepackage{mathrsfs}  
\usepackage{natbib}
\usepackage{amssymb}
\bibpunct{(}{)}{;}{a}{}{,} 
\usepackage[english]{babel}
\usepackage{longtable}
\usepackage[english]{babel}
\usepackage{caption}
\usepackage{url}
\usepackage{hyperref}

\authorrunning{Taddia et al.}
\titlerunning{iPTF15dtg: a double-peaked SN Ic}

\begin{document}

\title{iPTF15dtg: a double-peaked Type Ic Supernova from a massive progenitor}

\author{F. Taddia\inst{1}
\and C. Fremling\inst{1}
\and J. Sollerman\inst{1}
\and A. Corsi\inst{2}
\and A. Gal-Yam\inst{3}
\and E. Karamehmetoglu\inst{1} 
\and R. Lunnan\inst{4}
 \and B. Bue\inst{5}
\and M. Ergon\inst{1} 
\and M. Kasliwal \inst{6}
\and P.~M. Vreeswijk\inst{3}
 \and P.~R. Wozniak\inst{7}}
\institute{The Oskar Klein Centre, Department of Astronomy, Stockholm University, AlbaNova, 10691 Stockholm, Sweden.\\ \email{francesco.taddia@astro.su.se}
 \and Department of Physics, Texas Tech University, Box 41051, Lubbock, TX 79409-1051, USA.
\and Department of Particle Physics \& Astrophysics, Weizmann Institute of Science, Rehovot 76100, Israel.
  \and Astronomy Department, California Institute of Technology, Pasadena, California 91125, USA.
  \and Jet Propulsion Laboratory, California Institute of Technology, Pasadena, CA 91109, USA.
 \and Cahill Center for Astrophysics, California Institute of Technology, Pasadena, CA 91125, USA.
\and Los Alamos National Laboratory, MS D436, Los Alamos, NM 87545, USA.}

\date{Received; accepted}

\abstract
{Type Ic supernovae (SNe~Ic) arise from the core-collapse of H (and He) poor stars, which could be either single WR stars or lower-mass stars stripped of their envelope by a companion. Their light curves are radioactively powered and usually show a fast rise to peak ($\sim$10$-$15 d), without any early (first few days) emission bumps (with the exception of broad-lined SNe~Ic) as sometimes seen for other types of stripped-envelope SNe (e.g., Type IIb SN~1993J and Type Ib SN~2008D).}
{We have studied iPTF15dtg, a spectroscopically normal SN~Ic with an early excess in the optical light curves followed by a long ($\sim$30 d) rise to the main peak. It is the first spectroscopically-normal double-peaked SN~Ic observed. We aim to determine the properties of this explosion and of its progenitor star.}
{Optical photometry and spectroscopy of iPTF15dtg was obtained with multiple telescopes. The resulting light curves and spectral sequence are analyzed and modelled with hydrodynamical and analytical models, with particular focus on the early emission.}
{iPTF15dtg is a slow rising SN~Ic, similar to SN~2011bm. Hydrodynamical modelling of the bolometric properties reveals a large ejecta mass ($\sim$10~$M_{\odot}$) and strong $^{56}$Ni mixing. The luminous early emission can be reproduced if we account for the presence of an extended ($\gtrsim$500~$R_{\odot}$), low-mass ($\gtrsim$0.045~$M_{\odot}$) envelope around the progenitor star. Alternative scenarios for the early peak, such as the interaction with a companion, a shock-breakout (SBO) cooling tail from the progenitor surface, or a magnetar-driven SBO are not favored.}
{The large ejecta mass and the presence of H and He free extended material around the star suggest that the progenitor of iPTF15dtg was a massive ($\gtrsim35$~$M_{\odot}$) WR star suffering strong mass loss. }


\maketitle
\section{Introduction}
\label{sec:intro}
Stripped-envelope (SE) supernovae (SNe) stem from 
the core-collapse of massive stars whose outer layers 
were removed. SNe~IIb and Ib present little or no signatures
of hydrogen but their spectra are helium rich. The spectra of SNe~Ic are 
also helium-poor \citep[e.g.,][]{filippenko97}.

SNe~Ic exhibit fast-rising light curves ($\sim$2 weeks) which are powered by $^{56}$Ni at peak. The peak is followed by a relatively rapid decline, and the nebular phase starts at $\sim$2 months. Typical expansion velocities of normal SNe~Ic 
are on the order of $\lesssim$10000 km~s$^{-1}$. There are SNe~Ic with faster ejecta, which are sometimes associated with 
long-duration gamma-ray bursts (GRB; \citealp{woosley06}). These are known as broad-lined SNe~Ic (SNe~Ic-BL).

Well observed SNe~Ic are, among others, SN~1994I \citep{filippenko95},  SN~2004aw \citep{taubenberger06} and SN~2007gr \citep{valenti08}. 
\citet{drout11} presented a first multi-band sample of SNe~Ic, 
\citet{taddia15} presented the sample of SNe~Ic from SDSS and \citet{bianco14} and \citet{modjaz14} presented the light curves and the spectra of 64/73 SE SNe obtained at the Harvard-Smithsonian Center for Astrophysics (CfA). 
\citet{cano13}, \citet{lyman16}, and \citet{prentice16} have recently collected the known objects from the literature. 

 These investigations showed that normal SN~Ic ejecta are typically on the order of ${2-4}~M_{\odot}$, energies are a few 10$^{51}$ erg, and the ejected $^{56}$Ni masses are typically 
$\sim{0.15-0.2}~M_{\odot}$. 

An important exception is SN~2011bm \citep{valenti12}, which is characterized by very massive ejecta (7$-$17~$M_{\odot}$), as inferred from the modelling of its broad light curve. \citet{valenti12} suggested a 30$-$50~$M_{\odot}$ progenitor star for SN~2011bm, which is consistent with the idea
that single Wolf-Rayet (WR) stars with $M_{ZAMS}~>$~25--30~$M_{\odot}$ and a massive stellar wind produce SNe~Ic. 
The most massive stars often outshine all the other stars in a galaxy and are responsible for much of the heavy-element nucleosynthesis, so understanding the final fate of the most massive stars is essential.
However, the relatively small ejecta mass derived for most SNe~Ic discussed above  suggests that many of 
these SNe rather come from binary systems where a companion star has stripped the H and He envelopes from the SN progenitor star \citep{smartt09,eldridge13}.

To understand the nature of the progenitor stars of SE SNe, it is of importance to study the very early supernova emission (first few days, as in the case of PTF10vgv presented by \citealp{corsi12}). 
The detection of an early peak  
can bring important information on the radius of the progenitor star \citep{piro13}, but also on the outer structure of the star, the degree of $^{56}$Ni mixing, 
as well as on the presence of a companion or a dense circumstellar material (CSM). 

Some SE~SNe, such as SN~Ib~1999ex \citep{stritzinger02},  SN~Ib 2008D \citep{soderberg08,malesani09,modjaz09}, SN~Ib/IIb iPTF13bvn \citep{fremling16}, and many SNe~IIb (e.g., SN~1993J, \citealp{richmond94}; SN~2011dh, \citealp{arcavi11}, \citealp{ergon14}), do show such early emission. The Type Ib/c SN 2013ge showed an early peak but only in the ultra-violet (UV) light curves \citep{drout15}. For SN~2008D the early peak in the optical has been explained by different scenarios: as the result of the presence of $^{56}$Ni in the outermost layers or as the consequence of a modified density structure within the helium progenitor star \citep{bersten13}; as a shock-breakout \citep{rabinak11}; or as the effect of a jet in the explosion \citep{mazzali08}. 
A double peak was observed also for the peculiar SN~Ib 2005bf \citep[e.g.,][]{folatelli06}, although in this case it is characterized by a longer time scale ($\sim$2 weeks past explosion, consistent with the explosion of a normal SE~SN). A magnetar was invoked to explain the second peak and the late emission of for SN 2005bf \citep{maeda07}. 
For SNe~IIb, the shock breakout cooling tail for a relatively large progenitor radius (a few hundred solar radii) has been proposed as the powering mechanism of this early emission. This is the way to probe the radius of the progenitor stars for these SNe.
For normal SNe~Ic, we have hitherto not observed any early-time peak. 

Superluminous SNe (SLSNe)~Ic also show double-peaked light curves \citep{leloudas12,nicholl15,nicholl16}. SLSNe are very luminous ($M_R~<~-21$~mag) transients, probably arising from the explosion of very massive star \citep{quimby11,galyam12}.  In this case the first peak has been proposed to be powered by the post-shock cooling of extended stellar material \citep{nicholl15,piro15}, whereas the main peak could be powered by a magnetar. \citet{kasen15} suggested that also the first peak seen in SLSNe could be a signature of a central magnetar, whose energy inflates a bubble that drives a shock through the SN ejecta.

Finally, some SNe~Ic-BL connected to GRBs have also shown early emission due to the presence of a luminous GRB afterglow (e.g., SN 2013dx, \citealp{delia15}). SN~2006aj, a SN~Ic-BL associated with an X-ray flash \citep[e.g.,][]{sollerman06}, also showed a double peak. The early maximum was interpreted as due to the presence of an extended envelope formed by a dense wind \citep{campana06,waxman07,irwin15}. Similar to SN~2006aj, also the GRB-SN~2010bh exhibited an early peak \citep{cano11}. The shock break-out through an extended and low-mass envelope could power the early peak \citep{margutti15,nakar15_06aj}. 

In this paper, we present iPTF15dtg, the first spectroscopically normal SN~Ic with a detected double peak in the optical light curves. 
Through the analysis of the SN light curves, we estimate progenitor and explosion properties consistent with the core collapse of a 
massive WR star which suffered strong mass loss prior to its explosion.

The paper is structured as follows: In Sect.~\ref{sec:obs} we present the observations and how the data were reduced; in Sect.~\ref{sec:hg} we discuss the host galaxy. Section~\ref{sec:lc} shows the SN light curves, and Sect.~\ref{sec:spec} presents the SN spectra. In Sect.~\ref{sec:model} we analyze and model the SN data. The results from the analysis are discussed in Sect.~\ref{sec:discussion}, and our conclusions are given in Sect.~\ref{sec:conclusions}.

\section{Observations and data reduction}
\label{sec:obs}

\begin{figure}[t]
\centering
\includegraphics[width=9cm]{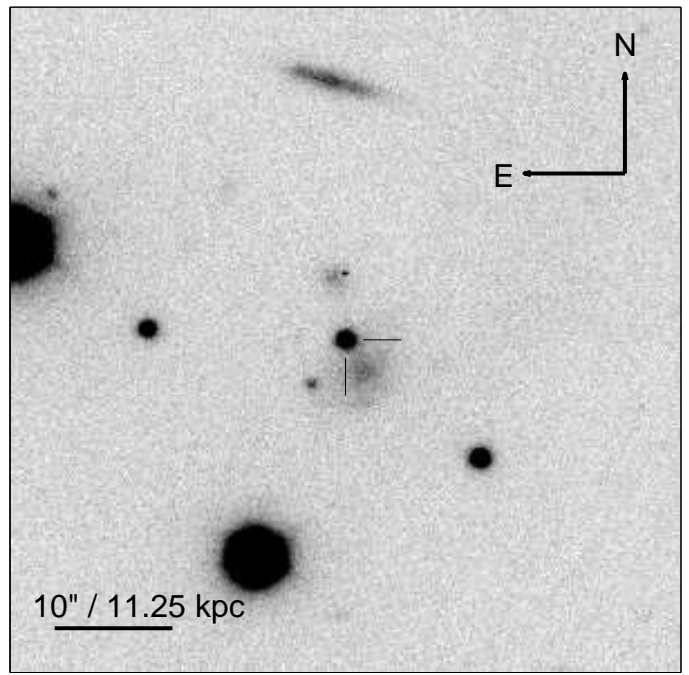}
\caption{\label{fc}iPTF15dtg (marked by two black segments) and its host galaxy in a $g-$band frame taken at the NOT on Jan. 10 2016 with ALFOSC. The orientation of the image is indicated in the top-right corner, whereas the scale is shown in the bottom-left.}
\end{figure}

Supernova iPTF15dtg was discovered on JD~2457333.931 at $g~=~19.63\pm$0.16~mag at RA~$=$~02:30:20.05 and DEC~$=$~+37:14:06.7 (J2000.0) on an image taken using the 48-inch Samuel Oschin telescope (P48) at Palomar Observatory, equipped with the 96~Mpixel mosaic camera CFH12K \citep{rahmer08}. The SN was not detected on JD~2457332.933 or JD~2457332.965 (i.e., 0.966 days before discovery) at limiting magnitude 
$g~\geq20.46$~mag and 20.16~mag, respectively. In the following we adopt the average between the epochs of last non-detection and discovery as the explosion date ($t_{\rm explo}=$JD~2457333.448$\pm$0.483), unless specified differently. Times are given with respect to this date in the observers frame 
throughout the paper, unless otherwise specified. 

We followed the SN with the P48 in $g$ band until $\sim$100~d after discovery. P48 photometry was reduced with the Palomar Transient Factory Image Differencing and Extraction (PTFIDE) pipeline\footnote{\href{http://spider.ipac.caltech.edu/staff/fmasci/home/miscscience/ptfide-v4.5.pdf}{http://spider.ipac.caltech.edu/staff/fmasci/home/miscscience/ptfide-v4.5.pdf}}$^{,}$\footnote{\href{ http://web.ipac.caltech.edu/staff/fmasci/home/miscscience/forcedphot.pdf}{ http://web.ipac.caltech.edu/staff/fmasci/home/miscscience/forcedphot.pdf}}, which performs template subtraction and PSF photometry. 

We also obtained photometry with the Palomar 60-inch telescope (P60; \citealp{cenko06}) in $Bgri$ bands, starting 2~d after discovery ($B-$band coverage started from $+$4.5~d). We furthermore used the Nordic Optical Telescope (NOT; \citealp{djupvik10}) to monitor the SN in $gri$ during the post-peak phase. The last images were obtained $\sim$130 days after explosion with the NOT. 
In Fig.~\ref{fc} we show iPTF15dtg in its host galaxy, in a $g-$band image taken on Jan. 10 2016 with the NOT. 
P60 and NOT photometric data were reduced using the pipeline presented in \citet{fremling16}. In Table~\ref{tab:phot} we report a log of the photometric observations. 
As reference stars to calibrate the P60 photometry, we used 14 stars in the SN field which were in turn calibrated using a Sloan Digital Sky Survey (SDSS; \citealp{ahn14}) field observed at similar airmass.
The final light curves are presented after combining the magnitudes obtained the same night.
 
Ten optical spectra were obtained from +3~d until +123~d, using the Telescopio Nazionale Galileo (TNG) $+$ DOLoRes, the Keck $+$ the Low Resolution Imaging Spectrometer (LRIS; \citealp{oke95}), the NOT $+$ the Andalusia Faint Object Spectrograph and Camera (ALFOSC), the Discovery Channel Telescope (DCT) $+$ the DeVeny spectrograph $+$ the Large Monolithic Imager (LMI), and the Gemini North telescope $+$ GMOS. The spectra were reduced in the standard manner, including wavelength calibration using an arc lamp, and flux calibration using a standard star (for each telescope we made use of dedicated pipelines, as in \citealp{fremling16}). In Table~\ref{tab:spectra} we report our spectral log. 

We also observed iPTF15dtg with the Karl G. Jansky Very Large Array \citep[VLA;][]{perley09} under our Target of Opportunity program\footnote{VLA/15A-314; PI: A. Corsi}. The first observation was carried out on 2015 December 17 ($+40~$d), between 00:28:40 and 01:28:30 UT, with the VLA in its D configuration. A second observation was carried out on 2016 January 7 ($+61~$d), between 03:35:58 and 04:35:41 UT, with the VLA in its DnC configuration. VLA data were reduced and imaged using the Common Astronomy Software Applications (CASA) package. Both observations yielded non-detections. We thus set the following $3\sigma$ upper-limits on iPTF15dtg radio flux: $\lesssim 23\,\mu$Jy at 6.4 GHz during the first epoch, and $\lesssim 20\,\mu$Jy at 6.2 GHz during the second epoch.

\section{Host galaxy}
\label{sec:hg}

iPTF15dtg was located in an anonymous galaxy at redshift $z~=~$0.0524$\pm$0.0002, which corresponds to a luminosity distance $D_L$~$=$~232.0~Mpc and distance modulus $\mu~=$36.83~mag. Here we assumed WMAP 5-years cosmological parameters \citep[$H_{0}$~$=$~70.5~km~s$^{-1}$~Mpc$^{-1}$, $\Omega_{M}$ $=$ 0.27, $\Omega_{M}$ $=$ 0.73,][]{komatsu09}. The redshift was determined from the gaussian fit of some of the host-galaxy emission lines (H$\alpha$, H$\beta$, [\ion{O}{iii}]~$\lambda$5007) superimposed on the SN spectra (see Sect.~\ref{sec:spec}). 

We assume that no host extinction affects the emission of iPTF15dtg, as we do not detect any narrow \ion{Na}{i}~D absorption lines at the host-galaxy rest wavelength. The Milky Way extinction in the $Bgri$ bands is $A_B~=$~0.235~mag, $A_g~=$~0.214~mag, $A_r~=$~0.148~mag, and $A_i~=$~0.110~mag (\citealp{sf11}).

The host galaxy has integrated magnitudes of $M_g~=~-17.4$~mag, $M_r~=~-17.8$~mag, $M_i~=~-17.9$~mag. This corresponds to a global metallicity of 12$+$log(O/H)~$=$~8.29 ($Z/Z_{\odot}~=~0.39$) following the luminosity-color-metallicity relation by \citet{sanders13}, or $Z/Z_{\odot}=0.32$ using the luminosity-metallicty relation by \citet{arcavi10}. From the SN spectrum taken on Dec. 6 2015, we could measure the emission line fluxes of the host-galaxy at the exact SN position. Based on their flux ratios, we derived a metallicity of 12$+$log(O/H)~$=$~8.22$\pm$0.20 ($Z/Z_{\odot}~=~0.34$) at the SN location using the O3N2 calibration by \citet{pp04}. The inferred metallicity at the location of iPTF15dtg is lower than that found for most of the other normal SNe~Ic \citep{sanders12} and more similar to the metallicities for SNe from blue supergiant stars, SLSNe, and SN impostors \citep{taddia13met,lunnan14,taddia15met}. 
The natal metallicity of iPTF15dtg appears to be comparable with that of long-duration GRBs 
 (\citealp{kruhler15} found a range of host metallicities 
 of 12$+$log(O/H) $=$ 7.9--9.0, with a median of 8.5).

\section{Light curves}
\label{sec:lc}

\begin{figure}
\centering
\includegraphics[width=9cm]{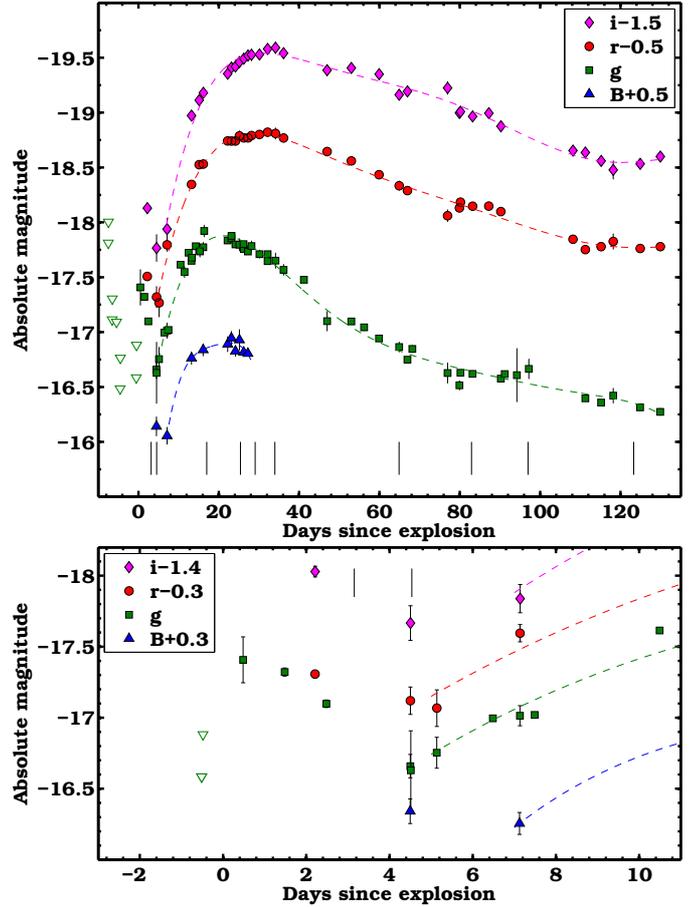}
\caption{\label{absmag}\textit{(Top panel)} $Bgri$ absolute-magnitude light curves of iPTF15dtg from P48, P60 and NOT. The main peak is fit by low-order polynomials, shown as dashed lines. The epochs of spectral observations are marked by vertical black segments. Pre-explosion magnitude limits in $g$ band are marked by triangles. In the $gri$ filters we notice the presence of an early peak, unprecented among spectroscopically normal SNe~Ic. 
\textit{(Bottom panel)} The early $Bgri$ light curves of iPTF15dtg.}
\end{figure}

\begin{figure}
\centering
\includegraphics[width=9cm]{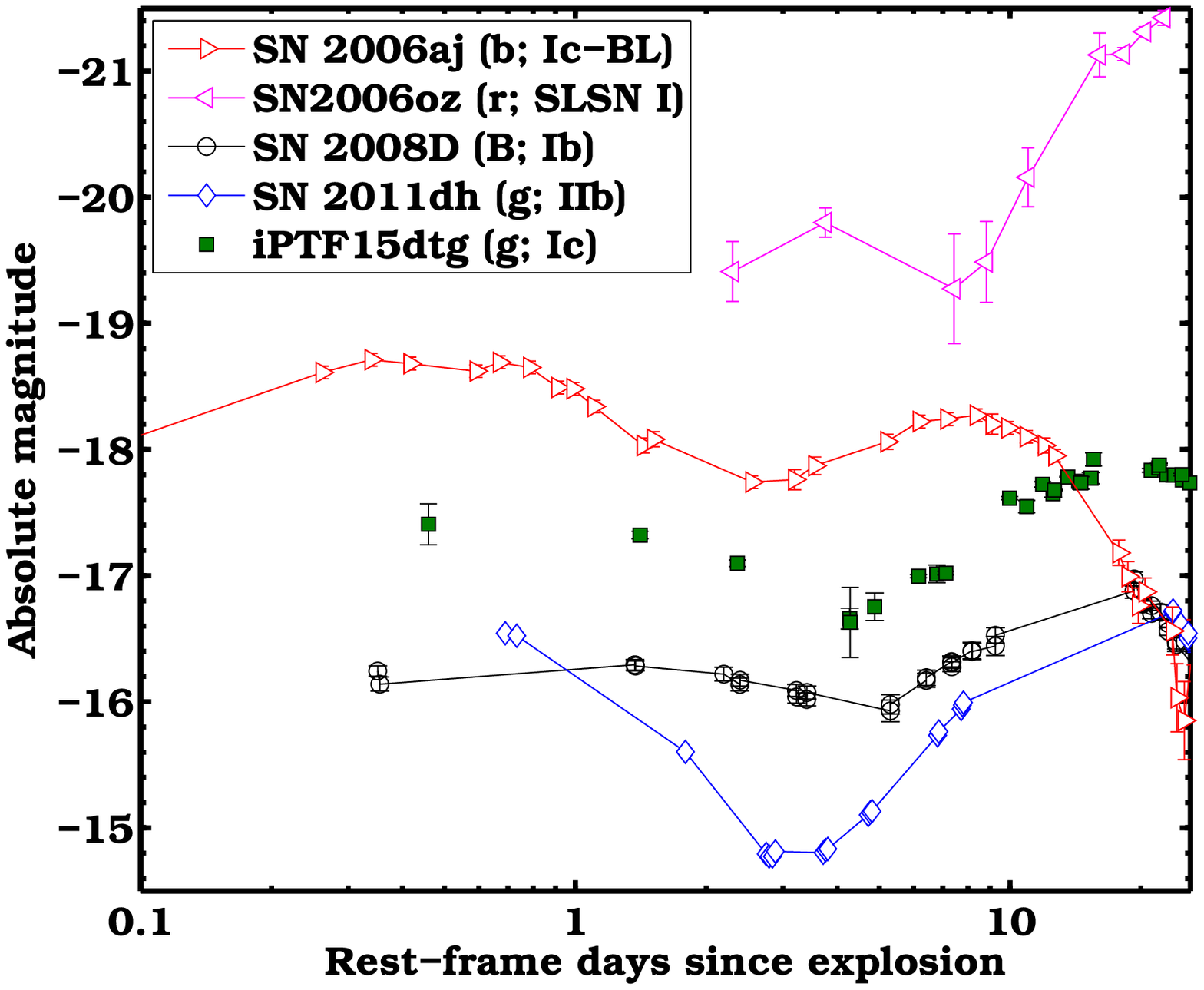}\\
\includegraphics[width=9cm]{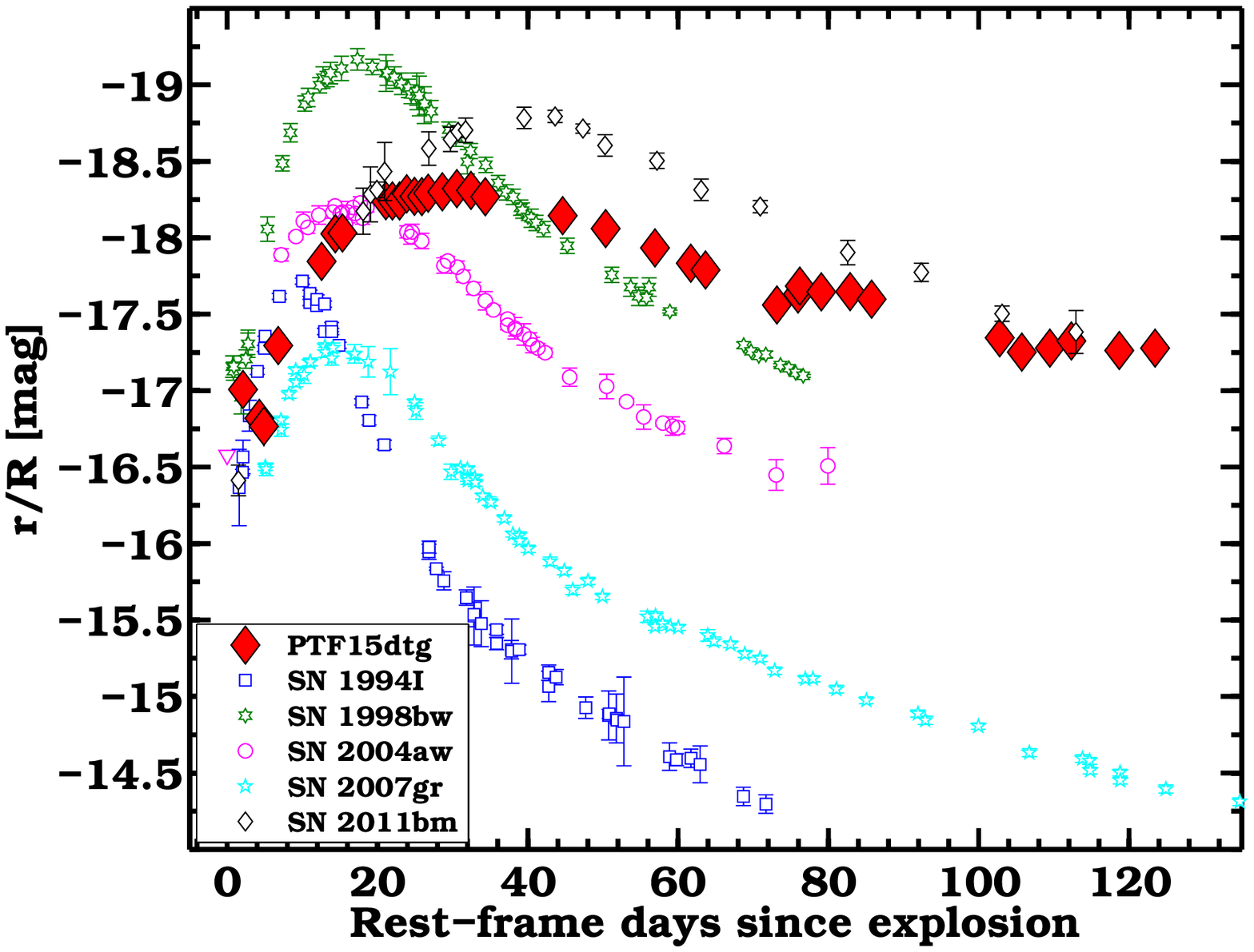}
\caption{\label{comp_abs}(Top panel) Comparison of the early $g-$band light curve of iPTF15dtg to the early light curves of other SE~SNe and SLSNe exhibiting a double peak. For each SN the filter and the SN type is reported in the legend. The data for SNe 2006aj, 2006oz, 2008D and 2011dh are taken from \citet{brown09}, \citet{leloudas12}, \citet{bianco14}, and \citet{arcavi11}, respectively. 
(Bottom panel) Absolute $r-$band magnitudes of iPTF15dtg compared to those of other well-studied SNe~Ic from the literature. Data for SNe~1994I, 1998bw, 2004aw, 2007gr, 2011bm are from \citet{richmond96}, \citet{clocchiatti11}, \citet{taubenberger06}, \citet{hunter09}, \citet{valenti12}, respectively.}
\end{figure}

\begin{figure}
\centering
\includegraphics[width=9cm]{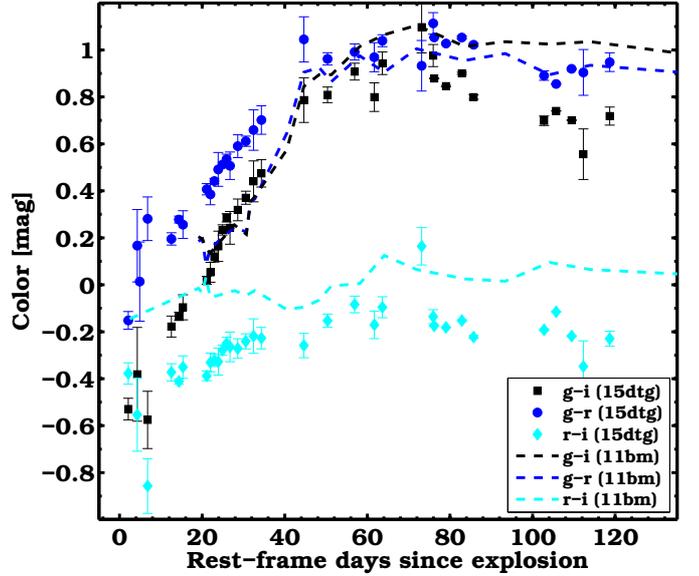}
\caption{\label{color}Color evolution of iPTF15dtg ($g-r$, $g-i$, $r-i$) compared to that of SN~2011bm \citep{valenti12}.}
\end{figure}

In Fig.~\ref{absmag} we present the $Bgri$ light curves
of iPTF15dtg. The $g-$band emission exhibits an early peak at $-$17.4 mag, followed by a declining phase lasting for about 4 days (see the bottom panel). Also the $r$ and $i$ bands show this early declining phase. In the $B$ band we detect only one epoch on the early declining phase, as the observations in this filter began a few days later. 

This early declining phase has been observed in other SE-SNe and in SLSNe, as shown in the top panel of Fig.~\ref{comp_abs}. Here we compare the absolute magnitudes of the early phase of iPTF15dtg to those of four other SNe belonging to different types and exhibiting a double-peaked light curve. The early peak of iPTF15dtg in $g$ band has a timescale similar to that of the other events, whereas its luminosity is in-between those of SN~Ic-BL~2006aj and SLSN~2006oz and those of SN~Ib 2008D and SN~IIb 2011dh.

After the first decline, 
all the light curves start to rise, reaching maximum (see top panel) at relatively late epochs for a SN~Ic, namely at $+$22.6, $+$21.2, $+$28.7 and $+$30.9 days in $B$, $g$, $r$, and $i$, respectively 
(21.5, 20.1, 27.3, 29.4 days in the rest frame). 
We determined the maximum epochs by fitting the light curves by low order polynomials, marked in Fig.~\ref{absmag} by dashed lines.

The light curves of iPTF15dtg are quite broad, and are characterized by $\Delta m_{15}~=~$0.34, 0.16, 0.11 ~mag  in $g$, $r$, $i$, respectively   (in the observer frame). The broadness of the light curves of iPTF15dtg is evident when 
we compare its $r-$band light curve with those of other well observed SNe~Ic (see the bottom panel of Fig.~\ref{comp_abs}). iPTF15dtg peaks later than any other SN~Ic, with the exception of SN~2011bm. Also SN~Ic~2010mb \citep{benami14} shows a late peak, but that SN is characterized by strong emission lines in the spectra due to the interaction with its CSM. 
On the other hand, iPTF15dtg 
 ($M_r^{max}~=~-$18.3~mag) 
is only slightly brighter than normal SNe~Ic such as SN~2007gr 
with a peak similar to that of SN~2004aw and slightly fainter than SN~2011bm, 
but clearly fainter than a SN~Ic-BL such as SN~1998bw. The $r-$band decline rate of iPTF15dtg at $\gtrsim+80$~d is slower 
than that of the other SNe~Ic, including SN~2011bm. 

The color evolution of iPTF15dtg is shown in Fig.~\ref{color}. 
The fast cooling in the early phase is clearly visible in $g-i$. Thereafter all the colors show a slower trend to the red, until $\sim$50 days when they are flatter. The color evolution of iPTF15dtg is very similar to that  of SN~2011bm. For iPTF15dtg $g-r$ is redder at pre-peak epochs (before $\sim$40~d), $g-i$ is bluer at later epochs, $r-i$ is bluer at all epochs.

\section{Spectra}
\label{sec:spec}

\begin{figure}
\centering
\includegraphics[width=9cm]{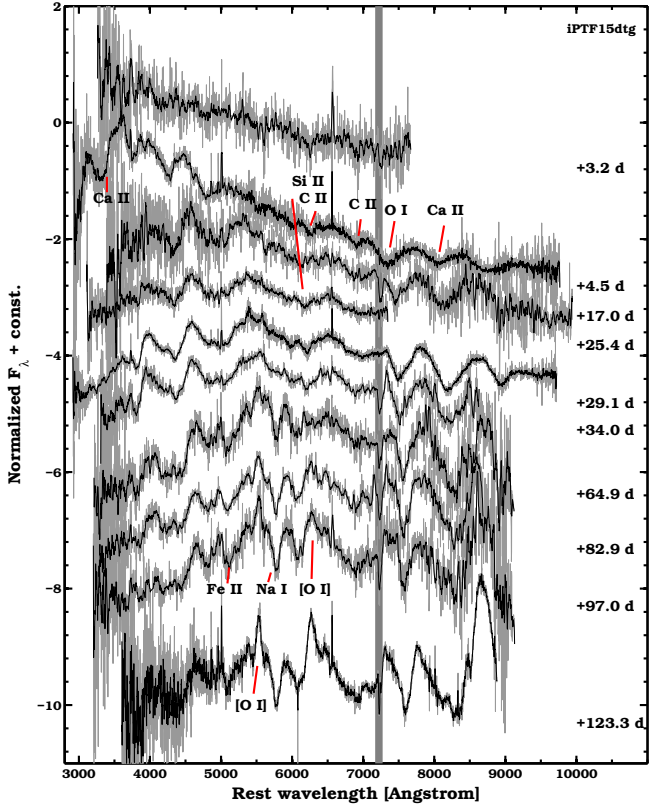}
\caption{\label{spec}Spectral sequence of iPTF15dtg. In gray we show the unbinned spectra, in black the same spectra after smoothing. The spectra are normalized by their median and shifted vertically for clarity. The spectra are de-redshifted but they are not corrected for extinction. For each spectrum we report its phase (in days since explosion). The main telluric feature is marked by a gray area. Line identifications are reported for some of the main features.}
\end{figure}

\begin{figure}
\centering
\includegraphics[width=9cm]{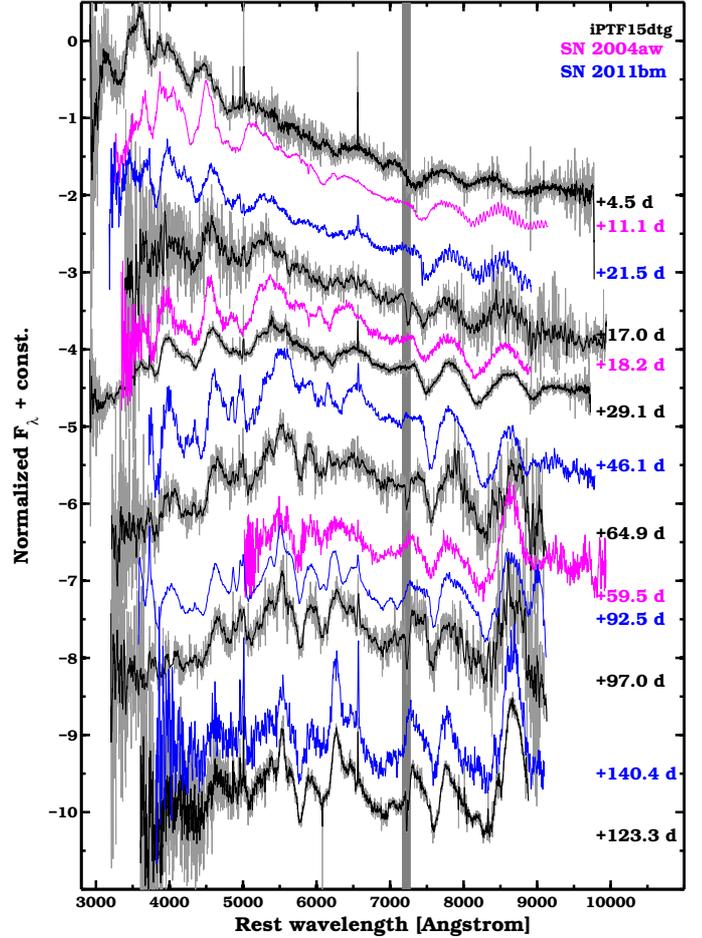}
\caption{\label{speccomp}Spectral comparison of iPTF15dtg to the SNe~Ic 2004aw \citep{taubenberger06} and 2011bm \citep{valenti12}. Each spectrum is corrected for redshift and extinction. Their phases are reported in days since explosion.}
\end{figure}

\begin{figure}
\centering
\includegraphics[width=9cm]{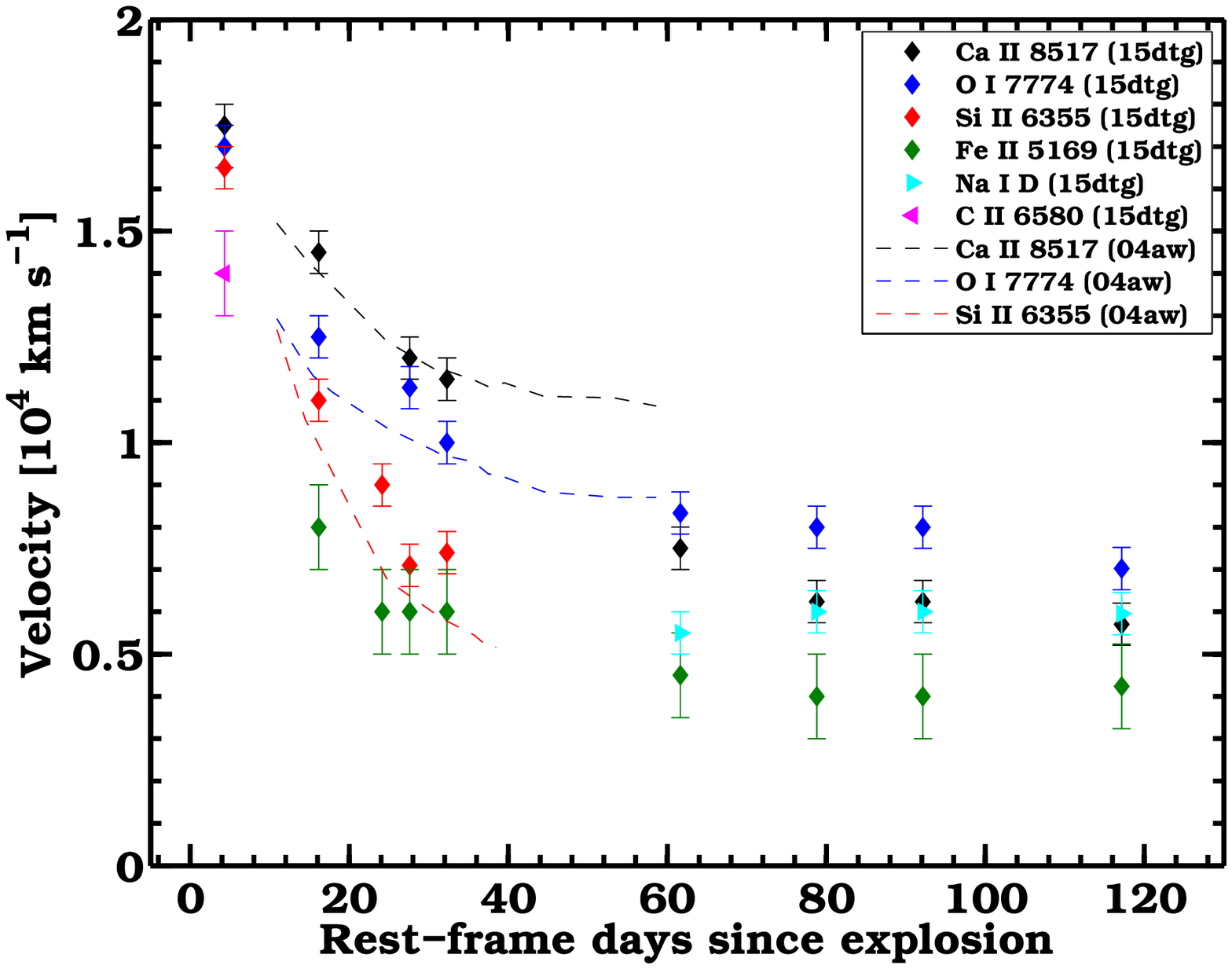}
\caption{\label{vel}P-Cygni minima velocities for different lines in the spectra of iPTF15dtg as compared to those of SN~2004aw from \citet{taubenberger06}.}
\end{figure}

iPTF15dtg was spectroscopically followed from +3~d until +123~d. The spectral sequence is shown in Fig.~\ref{spec}. All the spectra will be released via WISeREP \citep{yaron12}. 
The initially blue almost featureless spectrum is followed by a progressive reddening with the characteristic lines of a SN~Ic emerging from the continuum. No H or He lines are detected. The spectra are dominated by Ca, O and Fe lines. \ion{Ca}{ii} and \ion{O}{i} characterize the red part of the spectrum, with broad P-Cygni profiles. \ion{Fe}{ii}~$\lambda$5169 is visible after +17~d,  \ion{Na}{i}~D emerges at $\sim$+60 days and [\ion{O}{i}]~$\lambda$5577, [\ion{O}{i}]~$\lambda$6300 at $\sim$+80~d. In the first two spectra \ion{C}{ii} features (at 6580~\AA\ and 7234~\AA) are also identified. 

Figure~\ref{speccomp} illustrates that the spectra of iPTF15dtg resemble those of SN~2004aw, and also those of SN~2011bm. 
The spectra show an almost identical pattern in the red part, due to the P-Cygni profiles of \ion{Ca}{ii} and \ion{O}{i}. These three SNe have spectra that also show three emission peaks in the blue part, at $\sim$4000, 4800, 5500~\AA, as well as \ion{Si}{ii} below the narrow H$\alpha$ emission line from the host. 
The $+$123~d spectrum of iPTF15dtg is remarkably similar to that of SN~2011bm taken at a similar phase ($+$140~d), with both spectra exhibiting nebular features, in particular strong [\ion{O}{i}]  emission at 5577~\AA\ and 6300~\AA.

The spectral similarity between these SNe implies that also the velocity evolution of the different lines are quite similar, in particular between iPTF15dtg and SN~2004aw, as shown in Fig.~\ref{vel}. Here we compare the velocities of \ion{Ca}{ii}, \ion{O}{i}, and \ion{Si}{ii}, which we derived from the absorption minima of the P-Cygni profiles. We also report the velocities of \ion{Na}{i}~D and \ion{C}{ii}~$\lambda$6580 for iPTF15dtg.
 For iPTF15dtg we also measured the \ion{Fe}{ii}~$\lambda$5169 velocities, which are used to model the SN bolometric light curve (see Sect.~\ref{sec:model}). These measured velocities are consistent with those of normal SNe~Ic \citep{modjaz15}. All the line velocities decrease with time up to $\sim30~$d, thereafter they are almost constant.

\section{Modeling}
\label{sec:model}

\subsection{Bolometric properties}

\begin{figure}
\centering
\includegraphics[width=9cm]{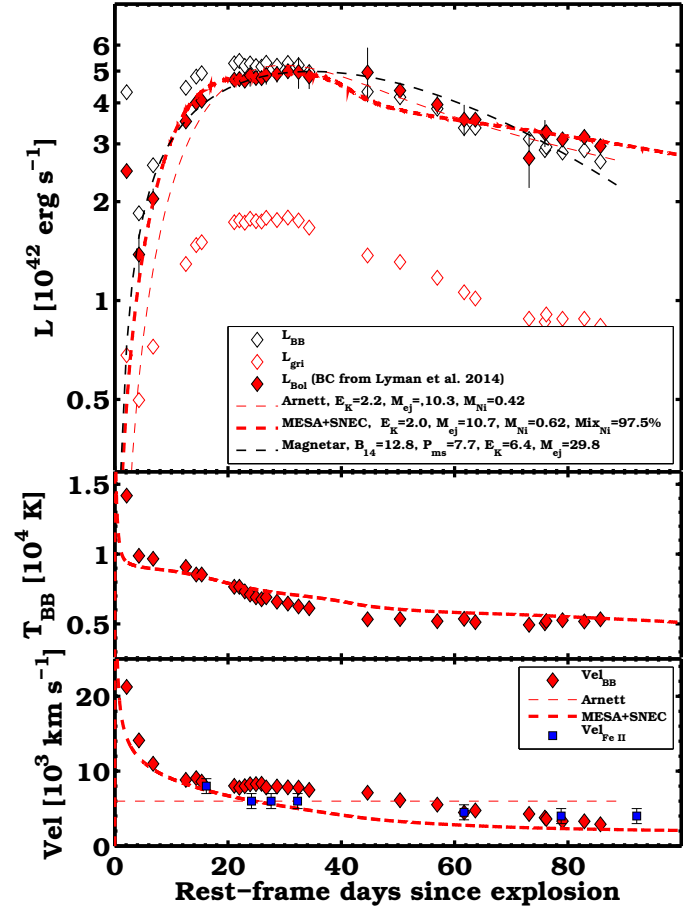}
\caption{\label{LTR}(Top panel) Bolometric light curves of iPTF15dtg.
The empty black diamonds represent the bolometric light curve obtained from the BB fit to the SEDs. The empty red diamonds represent the $gri$ integrated luminosity. The filled red diamonds 
mark the bolometric luminosity obtained using the bolometric corrections presented by \citet{lyman14}. The best Arnett fit to this light curve is shown by a red dashed line. The best hydrodynamical model fit is shown by a thick dashed line. The best magnetar model is shown by a black dashed line.
(Central panel) BB-temperature evolution for iPTF15dtg, along with the temperature evolution of the best hydrodynamical model. (Bottom panel) BB velocity (red diamonds) and \ion{Fe}{ii} velocities for iPTF15dtg. The best hydrodynamical model fit is shown by a thick dashed line, the velocity adopted for the Arnett model is represented by a thin dashed line, and fits the  \ion{Fe}{ii} velocities around peak.}
\end{figure}

Using the $gri$ light curves we build a quasi bolometric ($L_{gri}$) and a bolometric ($L_{BB}$) light curve, shown in the top panel of Fig.~\ref{LTR}. 
To obtain $L_{gri}$ we linearly interpolate the light curves at the same epochs, we convert the extinction-corrected $gri$ magnitudes into specific fluxes at the effective wavelength \citep{fukugita96} of their filters, and integrate the resulting spectral energy distribution (SED). To build $L_{BB}$ we fit a black-body (BB) function to the SED, in order to account also for the flux in the near-infrared (NIR) and in the UV. The fluxes obtained from the integrated SED and BB are multiplied by 4$\pi D_L^2$, where $D_L$ is the luminosity distance.  
We do not consider $B$ band in our SED because it has a limited temporal coverage and because it could be affected by line blanketing.

As our SEDs do not include emission in the UV or in the NIR, we check if $L_{BB}$ is consistent with the bolometric light curve ($L_{Bol}$) that we derive using the bolometric corrections (BCs) for SE SNe presented by \citet{lyman14}. We make use of the BCs to convert $M_g$ and extinction-corrected $g-r$ into $L_{Bol}$. For the first epoch we use the BC listed in table 4 of \citet{lyman14}, whereas for the remaining epochs we use those listed in their table 2.
The resulting  $L_{Bol}$ is shown in red diamonds in the top panel of Fig.~\ref{LTR}.  $L_{Bol}$  turns out to be fainter than $L_{BB}$,
probably due to the line blanketing in the bluer part of the spectrum which makes the actual SED less luminous than the best BB fit. Hereafter we use $L_{Bol}$ obtained from the BCs of \citet{lyman14} as our bolometric light curve.

The BB fit on the SEDs gives us an estimate of the temperature ($T$), shown in the central panel of Fig.~\ref{LTR}.
It is clear that the ejecta are progressively cooling down. The cooling is particularly strong between the first and the second epoch, with $T$ dropping from 1.5$\times$10$^4$~K to 1.0$\times$10$^4$~K in $\sim$2 days.

In the bottom panel of Fig.~\ref{LTR} we show the BB-velocity (${\rm Vel}_{BB}$) as derived from the fit by dividing the resulting BB-radius by the time since explosion. The photosphere recedes in the slowest part of the ejecta as the time increases. 

\subsection{Modeling of the main peak}
\label{sec:modelmain}

\subsubsection{Arnett model}
\label{sec:arnett}
In order to derive estimates of the ejecta mass ($M_{ej}$), the $^{56}$Ni mass, and the explosion energy ($E_K$) for iPTF15dtg, we can fit the main peak of the bolometric light curve and the photospheric velocity using a simple Arnett model (\citealp[see e.g.,][]{valenti08,cano13,taddia15}). 
The model assumes 
constant opacity $\kappa$, which we set to 0.07~cm$^2$~g$^{-1}$. This value was also used by \citet{cano13} and  \citet{taddia15}, as it is appropriate for the electron scattering in H-poor SNe (the assumption of a constant opacity is obviously a limitation of this model, see e.g., \citealp{wheeler15} and \citealp{dessart16}). The model also assumes that the $^{56}$Ni is located at the center of the ejecta. We adopt the relation between $E_K$, $M_{ej}$ and the photospheric velocity that is valid for a sphere of constant density, as in \citet{taddia15}. 
We use 6000 km~s$^{-1}$ as the photospheric velocity in the Arnett model, from the measured \ion{Fe}{ii}~$\lambda$5169 P-Cygni minima (these values are shown as blue squares in the bottom panel of Fig.~\ref{LTR}).

By fitting the model to the bolometric light curve we obtain  $E_{K}~=~(2.2\pm0.1)\times$10$^{51}$~erg, $M_{ej}~=~10.3\pm0.6$~$M_{\odot}$, and M$(^{56}\rm Ni)~=~0.42\pm0.01$~$M_{\odot}$.  The given uncertainties only include the errors of the fit. The ejecta mass thus obtained is substantially higher than that of normal SNe~Ic.  The $^{56}$Ni mass to ejecta mass ratio ($\sim$0.04) is similar to that
of other normal SNe~Ic \citep{cano13}, and provides a good argument for radioactive heating (and against magnetar heating).  
The model, shown by a thin dashed line in the top panel of Fig.~\ref{LTR}, is overall able to reproduce the light curve with good accuracy, but it fails to provide a good fit at early epochs. The rise occurs later in the model, as a result of the centralized $^{56}$Ni distribution.

\subsubsection{Hydrodynamical model}
\label{sec:hydro}

\begin{figure}
\centering
$\begin{array}{c}
\includegraphics[width=9cm]{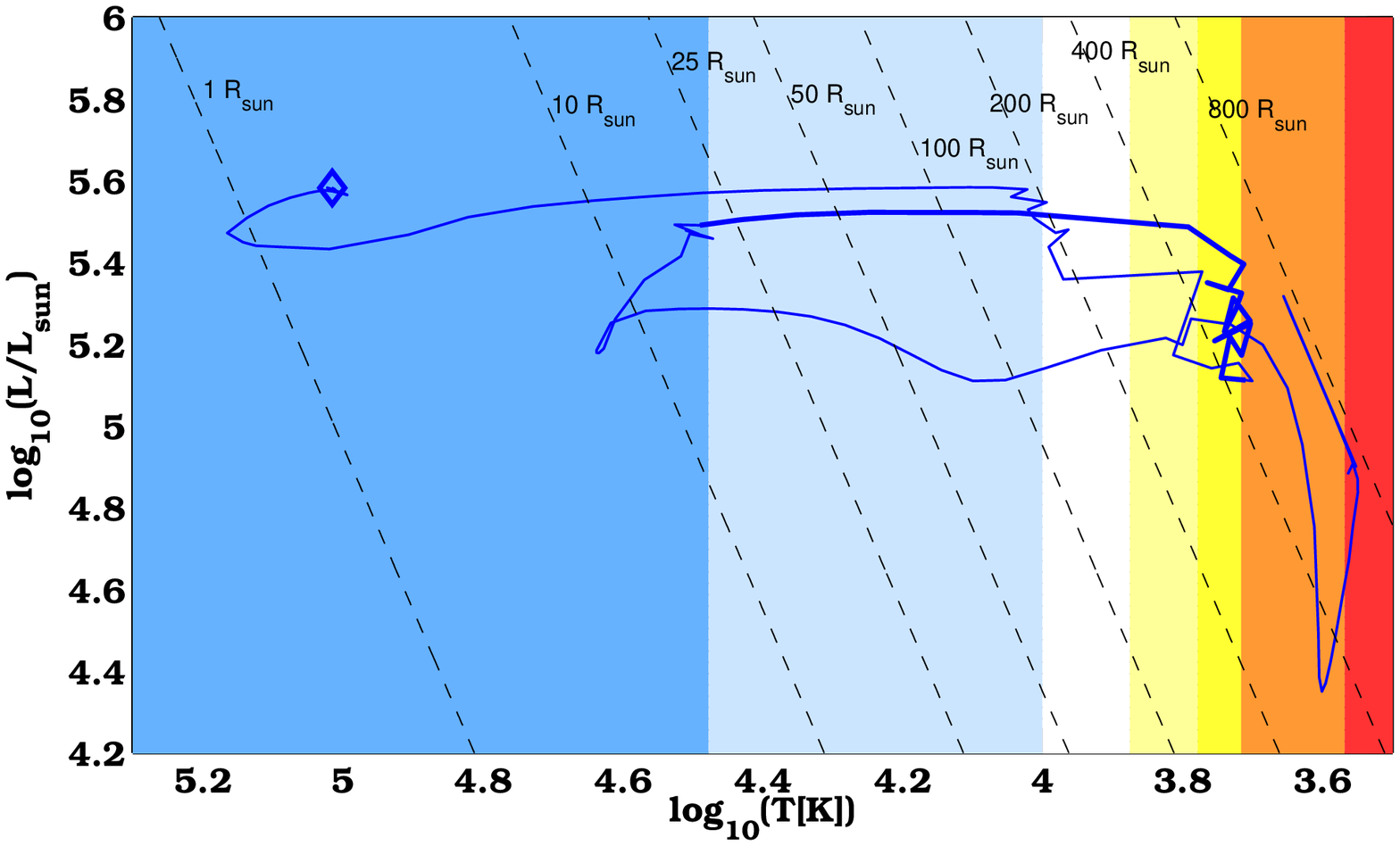}\\
\includegraphics[width=9cm]{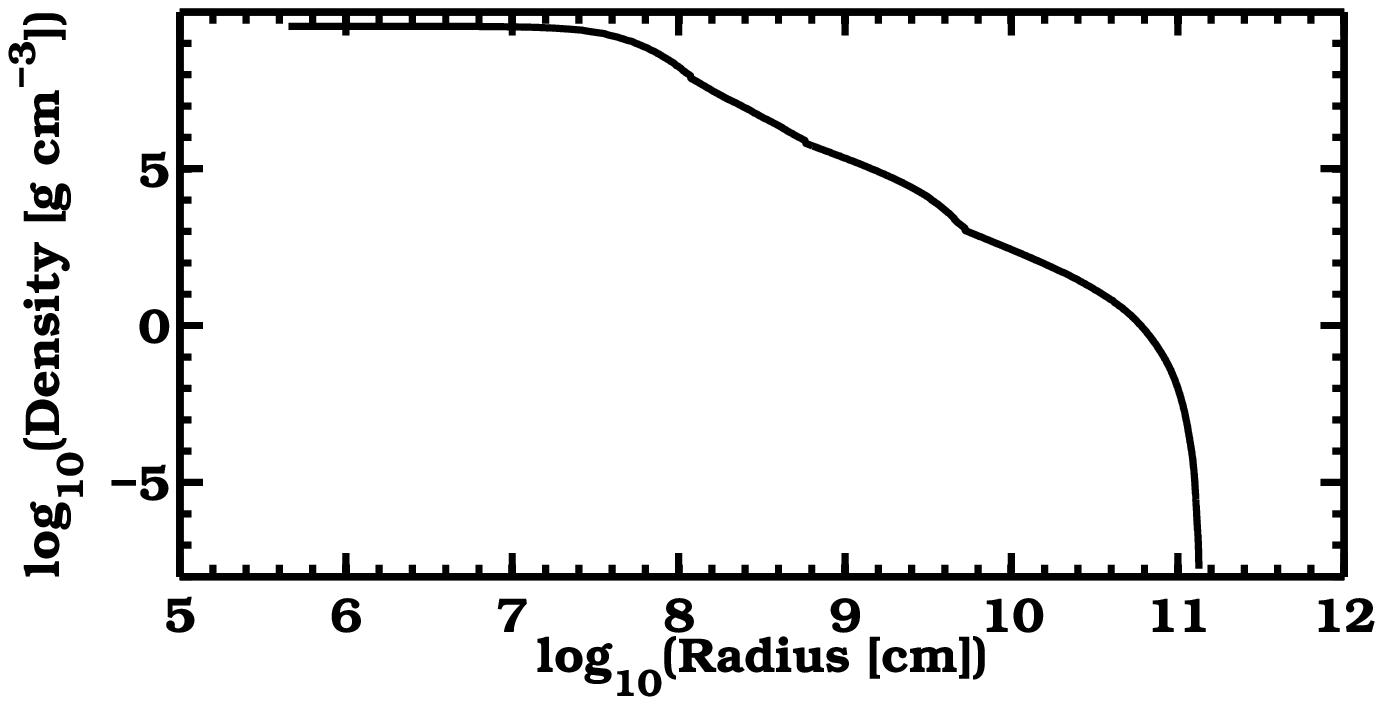}
\end{array}$
\caption{\label{HR} (Top panel) Evolution of our MESA star model in the HR diagram. The position of the progenitor star just before collapse is marked by a blue diamond. This progenitor star has a compact radius (see dashed black lines) and it is characterized by a rather high temperature (10$^{5}$ K), therefore it belongs to the O stellar type. The phase when most of the H mass is lost due to strong winds is marked by a thicker blue line. (Bottom panel)  Density profile of our progenitor star model.}
\end{figure}

Given the simplified assumptions in the Arnett model, we try to refine the estimates of the progenitor properties by making use of the Modules for Experiments in Stellar Astrophysics (MESA; \citealp{paxton11}) and of the SuperNova Explosion Code (SNEC; \citealp{morozova15}). MESA allows us to build a progenitor star via its stellar evolution code, whereas with the 1D hydrodynamical code SNEC we can explode this progenitor and calculate a bolometric light curve. 

Our approach is to construct a real stellar evolution and hydrodynamical model based on the input we have from the Arnett model fit to the SN data, without covering a large parameter space. We used MESA to obtain a star with a final mass of $\sim$12~$M_{\odot}$, which implies an ejecta mass of $\sim$10~$M_{\odot}$, as suggested by our Arnett model. We first set the metallicity of our stellar evolution model to $Z~=~0.00676~=~0.34~Z_{\odot}$, consistent with the metallicity measured 
from the host-galaxy emission lines
in Sect.~\ref{sec:hg}. 
Since we wanted to obtain a H-free progenitor (iPTF15dtg is a SN~Ic), we focused on single stars with initial masses $>30$~$M_{\odot}$, which have the necessary mass-loss rate to expel the outer H layer. We also considered that massive stars can be characterized by rapid rotation.  Given these boundary conditions, after running a series of models with MESA, we finally produced  a 12.1~$M_{\odot}$ H-free star, with a radius of $\sim$1.9~$R_{\odot}$, and a surface temperature of 10$^5$~K, that is consistent with a massive WR star. This star has a total helium mass of 4.96~$M_{\odot}$. 
To produce this result we evolved a star with an initial mass of $M_{\rm ZAMS}~=~35~M_{\odot}$, rotating at $v_{surf}$~$=$~105~km~s$^{-1}$. The evolution in the Hertzsprung-Russell  (HR) diagram of this star up to collapse is shown in the top panel of Fig.~\ref{HR}. The density profile of the progenitor star is shown in the bottom panel. Most of the H envelope is lost due to strong winds (the mass-loss rate reaches a peak of $\sim$1.3$\times$10$^{-1}~M_{\odot}~$yr$^{-1}$) when the star is about 5.3$\times$10$^{6}$ yr old.

We exploded this 12.1~$M_{\odot}$ H-free star leaving 10.7~$M_{\odot}$ in the ejecta, with an explosion energy of 2$\times$10$^{51}$~erg (as inferred from the Arnett model). These values allowed us to reproduce the  \ion{Fe}{ii}$~\lambda$5169 velocity profile of iPTF15dtg (see bottom panel of Fig.~\ref{LTR}).

\begin{figure}
\centering
\includegraphics[width=9cm]{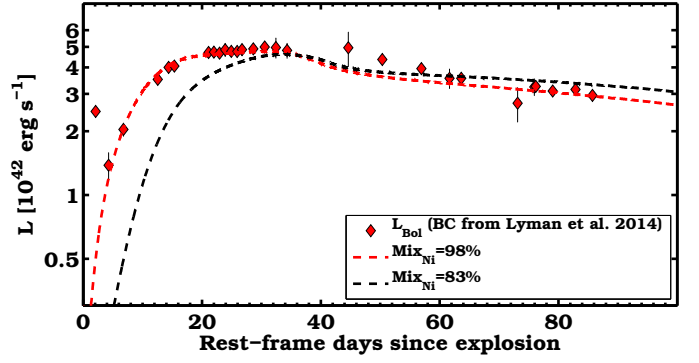}
\caption{\label{mix} The effect of $^{56}$Ni mixing on the light curve models of iPTF15dtg. Almost full mixing (red dashed line) is required to fit the early rise and the main peak shape. Even a model with a value of 83\% for $^{56}$Ni mixing (black dashed line) does not fit the early rise.}
\end{figure}

As the Arnett model does not assume any $^{56}$Ni mixing, and thus fails to reproduce the early rise to the main peak of the light curve, we resorted to
uniformly distribute the $^{56}$Ni up to the outer layers of the SN. After a few attempts, we found that by mixing the $^{56}$Ni out to 97.5\% of the progenitor mass, the hydrodynamical model nicely fits the rise and the flat peak of the bolometric light curve, as well as the declining phase. In Fig.~\ref{mix} we show how a lower degree of mixing (83\%) fail to reproduce the early rise and the peak shape.

We chose a $^{56}$Ni mass of 0.62~$M_{\odot}$. This was higher than the estimate from the Arnett model (this difference is mainly due to the mixing) but it allowed us to fit the SN peak with the hydrodynamical model. 
 This is shown in Fig.~\ref{LTR} (top-panel). The MESA+SNEC model also fits the temperature evolution, as shown in the central  panel of  Fig.~\ref{LTR}. 

Neither the best hydrodynamical model nor the Arnett model can reproduce the early peak.  In Sect.~\ref{sec:modelearly}, we make use of the parameters obtained for the ejecta mass and the explosion energy as input for other models which include the physics needed to explain the luminous early peak.

\subsubsection{Magnetar model}
\label{sec:magnetar}
There is also the possibility that the main peak is powered by a magnetar (see e.g., \citealp{maeda07} for SN~2005bf). This mechanism has been suggested for superluminous supernovae \citep{kasen10_mag}. We fit the magnetar model by \citet{kasen10_mag} (see also \citealp{inserra13}) to the bolometric light curve and to the \ion{Fe}{ii} velocity at the epoch of maximum, obtaining $E_{K}~=~(6.4\pm0.5)\times$10$^{51}$~erg, $M_{ej}~=~29.8\pm2.2$~$M_{\odot}$, B~$=$~(12.8$\pm$0.4)$\times~$10$^{14}$~G, and $P~=~7.7\pm0.6$~ms. Here $B$ is the magnetic flux density, $P$ is the rotation period of the magnetar. 
The model does reproduce the light curve (black dashed line in Fig.~\ref{LTR}), but not as well as the hydrodynamical radioactively-powered model, especially on the declining phase. Moreover, this SN is not superluminous and its spectra do not show features associated with the presence of a magnetar, therefore we disfavor this scenario.

\subsection{Modelling of the early peak}
\label{sec:modelearly}
The early ($\sim3$~d) peak in the light curves of iPTF15dtg is the first one ever observed for a spectroscopically normal SN~Ic. 
There are different scenarios that can potentially explain the presence of this feature, and in the following we discuss each of them.

Possible mechanisms are: 1) the shock breakout cooling (SBO) tail \citep[e.g.,][]{piro13}; 2) the emission from interaction with a companion star \citep[][]{kasen10}; 3) a magnetar-driven SBO \citep{kasen15}; 4) the emission from a progenitor star surrounded by an extended envelope of relatively small mass \citep[][]{nakar14,piro15,margutti15,nakar15_06aj}.

Each of these models depends on several parameters whose degeneracy is impossible to break by just fitting the early peak.
Therefore, we make use of the results derived from the hydrodynamical modelling of the main peak and fix some of the parameters to the values derived there (in Sect.~\ref{sec:discussion} we discuss some caveats related to the hydrodynamical modelling).
In particular we adopt the explosion energy and ejecta mass from Sect.~\ref{sec:hydro}. Whereas this approach does not provide a fully consistent model of the entire bolometric light curve, it allows us to reduce the degree of freedom and provide good estimates on the parameters that only the early emission can constrain.
The modeling in the following sections demonstrates that the only favourable scenario is the extended-envelope model.

\begin{figure}
\centering
\includegraphics[width=9cm]{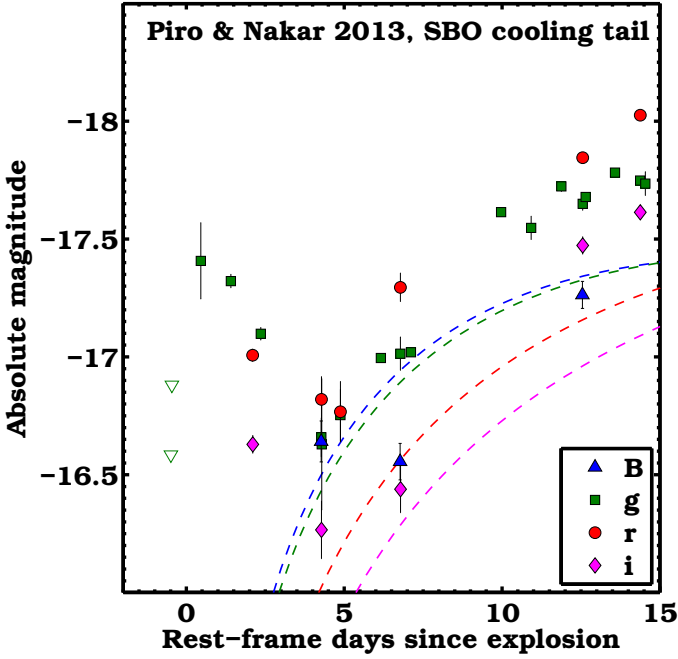}
\caption{\label{sbo}\citet{piro13} model (dashed lines) for the early light curve of iPTF15dtg, in $Bgri$ bands. In the modelling, we adopt the ejecta mass and explosion energy derived from the hydrodynamical modelling. The large ejecta mass implies a long time-scale for the cooling tail in the optical, which cannot reproduce the early peak of iPTF15dtg.}
\end{figure}

\begin{figure}
\centering
\includegraphics[width=9cm]{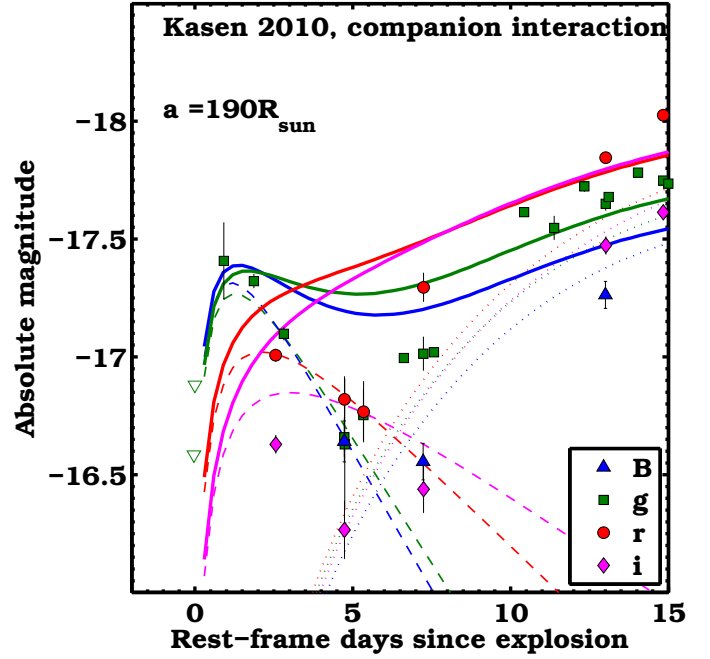}
\caption{\label{kasen}\citet{kasen10} model (dashed lines) for the early light curve of iPTF15dtg, in $Bgri$ bands. This model decribes the emission from the interaction with a companion star. In the modelling, we adopt the ejecta mass and explosion energy derived from the hydrodynamical modelling. We modified the original \citet{kasen10}, optimized for SNe~Ia, in order to reproduce the interaction of a compact WR star with a companion. 
Shown by dotted lines is the $Bgri$ emission from the $^{56}$Ni contribution, as derived from the hydrodynamical modelling. The \citet{kasen10} model alone reproduces the $B$, $g$ and $r$ band emission up to $+$5~d. However, if we also add the $^{56}$Ni contribution to the model, then there is a mismatch to the data due to the long time scale of the companion interaction process.}
\end{figure}

\begin{figure}[!h]
\centering
$\begin{array}{c}
\includegraphics[width=9cm]{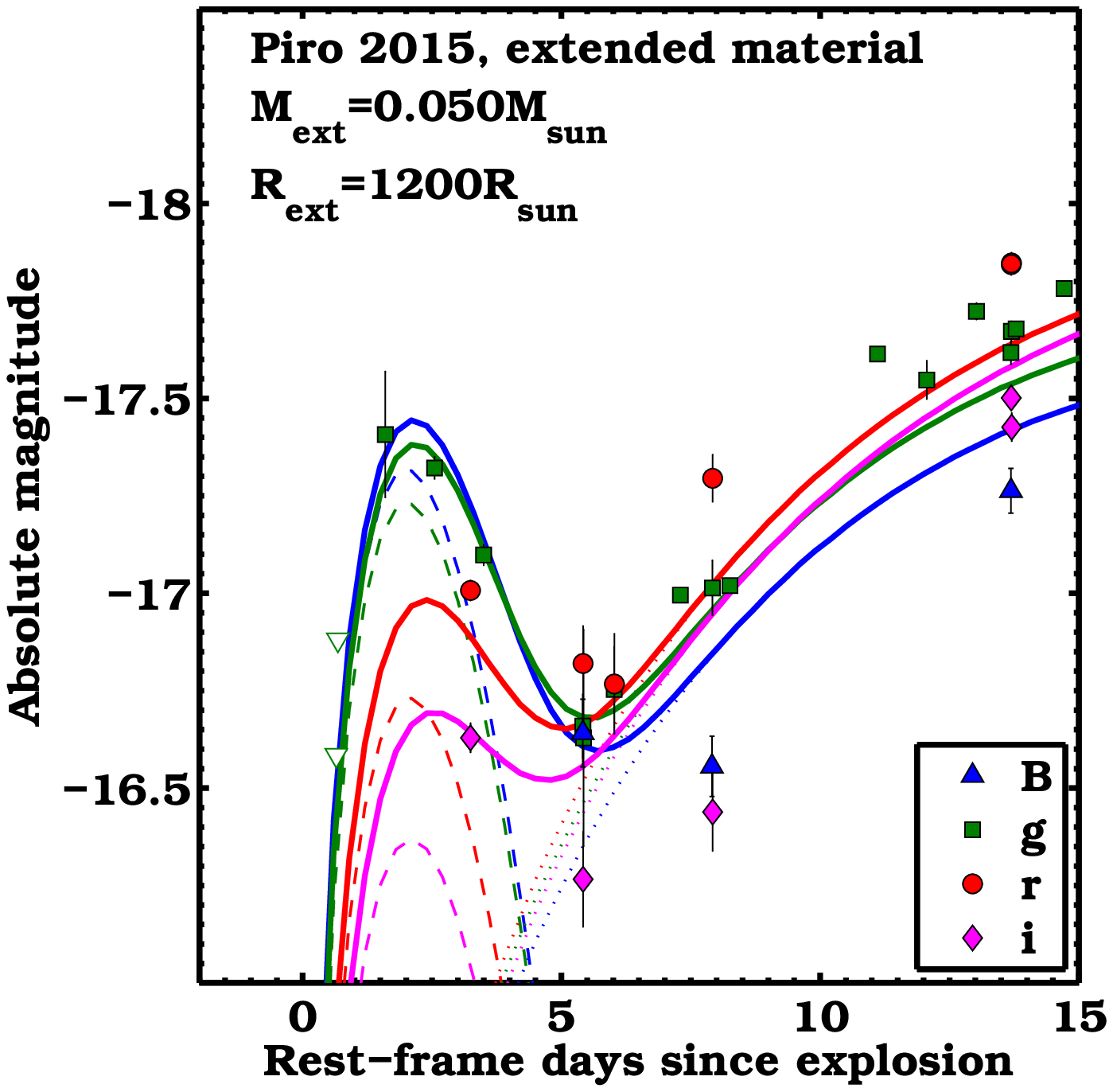}\\
\includegraphics[width=9cm]{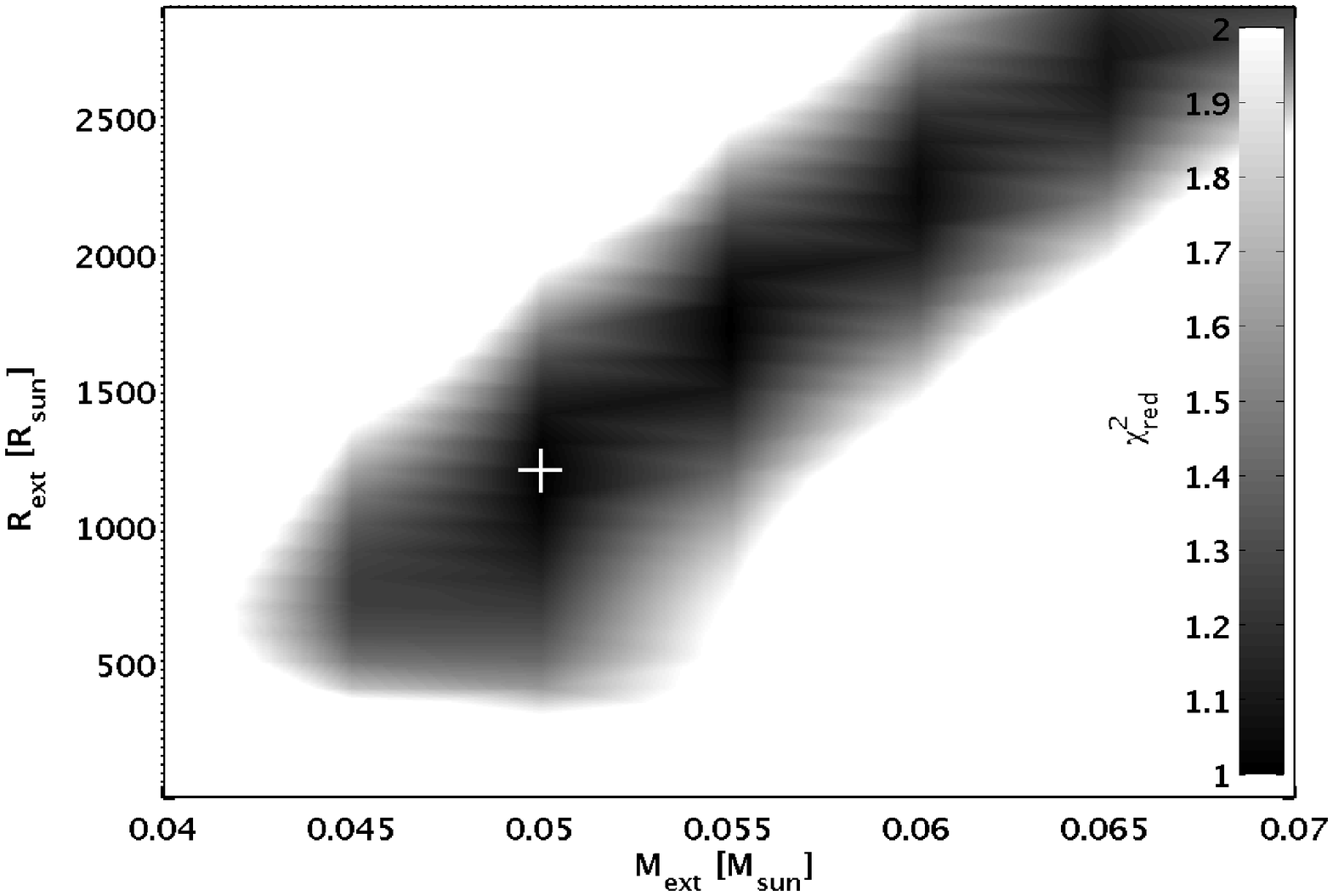}
\end{array}$
\caption{\label{nakar}\textit{(Top-panel)} \citet{piro15} model (dashed lines) for the early light curve of iPTF15dtg, in $Bgri$ bands. This model describes the emission from a star surrounded by extended material and it has been constructed to reproduce double-peaked SNe. In the modelling, we adopt the ejecta mass and explosion energy derived from the hydrodynamical modelling. Shown by dotted lines is the $Bgri$ emission from the $^{56}$Ni contribution, as derived from the hydrodynamical modelling. The sum of the \citet{piro15} model with the $^{56}$Ni contribution provides a good match to the early peak as well as to the rise to the main peak. This is due to the short timescale of the emission from the extended envelope. \textit{(Bottom-panel)} Reduced $\chi^2$ surface plot for the \citet{piro15} model fit, given different values of $R_{ext}$ and $M_{ext}$, for the explosion epoch derived from the best fit. The best $R_{ext}$ and $M_{ext}$ fit values are marked by a white cross. There is a clear degeneracy in the model fit for $R_{ext}\gtrsim~500~R_{\odot}$,  and $M_{ext}\gtrsim0.045$.}
\end{figure}

\subsubsection{The shock-break-out cooling-tail scenario}
The SBO scenario seems inconsistent with a small progenitor radius, like the one of the progenitor star we exploded to fit the main peak. Indeed in Fig.~\ref{LTR} the early peak in the bolometric light curve is not fit by the best hydrodynamical model. We use the analytic model from \citet{piro13} to further test the SBO cooling tail model. In their eqs. 1 and 2 \citet{piro13}
provide expressions for the luminosity and the temperature as a function of time. Assuming BB emission, we can derive the luminosity in the $Bgri$ bands to directly compare with the observed data.  In the model by \citet{piro13}, we use the explosion energy and the ejecta mass obtained in Sect.~\ref{sec:modelmain} from the hydrodynamical model of the bolometric light curve ($M_{ej}~=~10.7~M_{\odot}$, $E_{K}=2.0\times 10^{51}$~erg).
We also adopt $\kappa=0.2$~cm$^2$~g$^{-1}$, as we assume that the H-free gas is fully ionized at this stage. The explosion epoch is again assumed to be the average of last non-detection epoch and discovery epoch (the uncertainty is only $\sim$0.5~d).  Then, we solve for the only variable that is left, which is the progenitor radius. It turns out that it is not possible to provide a good fit of the early emission as the large ejecta mass implies a time scale for the cooling that is longer than the observed one. If we match the early peak $g-$band magnitude, we would obtain a large radius of $\sim$500 $R_{\odot}$, but the model clearly does not fit  (see Fig.~\ref{sbo}, where the SBO model in $Bgri$ bands for $R~=~510~R_{\odot}$ is marked by dashed lines). We therefore exclude a SBO cooling tail model for iPTF15dtg.

\subsubsection{The companion-interaction scenario}

In this section we try to test if the early emission can arise from the interaction between the SN ejecta and a putative companion star, as  described in the model by \citet{kasen10}. We set the inner and the outer ejecta density profiles to be power laws of the radius ($\rho\propto r^{-\delta}$ and $\rho\propto r^{-n}$) with $\delta=0$ and $n=6$, which is appropriate for a compact progenitor star \citep{chevalier82}. 
As in the SBO cooling-tail model we adopt $M_{ej}~=~10.7~M_{\odot}$, $E_{K}~=~2.0\times 10^{51}$~erg and $\kappa~=~0.2$~cm$^2$~g$^{-1}$.
The free parameters left are the binary separation, $a$, and the explosion epoch (which is however constrained by the good pre-discovery limits). If we fit the first 5 days in the light curves neglecting the $^{56}$Ni contribution, then the \citet{kasen10} model provides a good fit to at least the $B$, $g$ and the $r$ band, as shown by the dashed lines in Fig.~\ref{kasen}. The $i-$band flux is overestimated. The best fit is obtained for a binary separation of 190 $R_{\odot}$, and for an explosion epoch coincident with the last non-detection. We note that for larger values of the index $n$, the quality of the fit is worse. 
However, the $^{56}$Ni  contribution is important also in the first days as our best hydrodynamical model shows that the $^{56}$Ni is mixed out into the outer layers. A strong $^{56}$Ni mixing allowed us to fit the rise to the main peak and its flat shape. In Fig.~\ref{kasen} the emission from the $^{56}$Ni as derived from the hydrodynamical model (assuming BB emission) is represented by dotted lines in the $Bgri$ bands. If we sum the companion interaction emission and the $^{56}$Ni contribution, the total emission of the model (solid lines) is clearly in excess compared to the data in all the bands already at 3 days. 
This model might still work if we allow for a lower degree of $^{56}$Ni mixing. If we distribute the $^{56}$Ni up to 83\% of the progenitor mass instead of 97\%, we can get a good fit for the first 5 days in the $B$ and $g$ bands (and a reasonable fit to the $r$ band, but not in the $i$ band). However, we would strongly underestimate the flux in all the bands on the rise to the main peak. A strong $^{56}$Ni mixing thus seems to be necessary to fit the bolometric light curve rise, and makes it difficult to find a good fit for the \citet{kasen10} model. The transition between the first peak and the rise to the main peak is clearly sharper in the observed data than in the model shown by the solid lines in Fig.~\ref{kasen}, and thus the mechanism powering the main peak must have a timescale even shorter than the one of the \citet{kasen10} model, in all the bands. This is true also for other values of binary separation. 

We also stress that the last spectrum of iPTF15dtg reveal a strong [\ion{O}{i}]~$\lambda$6300 line that is similar to that of SN~2011bm, for which a massive, and thus probably single progenitor system was suggested \citep{valenti12}.

\subsubsection{The magnetar-driven shock-break-out cooling-tail}
\citet{kasen15} have recently shown that a magnetar-driven SBO could produce an early peak in a magnetar powered SN.  If we use the magnetar parameters from Sect.~\ref{sec:modelmain}, and modify the energy injected by the magnetar in the SN ejecta using eqs. 26 and 27 in \citet{kasen15}, we cannot simultaneously fit the time and the luminosity of the early peak of iPTF15dtg. Adopting a large (6$\times$10$^{52}$ erg) magnetar energy we can fit the epoch of the early peak, but the luminosity of this model turns out to be almost ten times brighter as compared to the data. Also for this reason, we do not favor the magnetar scenario as the main powering source of iPTF15dtg. 

\subsubsection{The extended-envelope scenario}
We can obtain a good fit to the early emission for all the bands using the double-peaked model by \citet{piro15}, which assumes that the progenitor star is surrounded by a low-mass, extended envelope. This could be in equilibrium, or mass ejected by the star during its life. In Fig.~\ref{nakar} the best fit is shown with colored solid lines, given by the sum of the \citet{piro15} model (dashed lines) plus the $^{56}$Ni contribution (dotted lines). For the \citet{piro15} model we have three independent free parameters, namely the extension of the low mass envelope ($R_{ext}$), its mass ($M_{ext}$), and the explosion epoch.  We assumed the same opacity, ejecta mass and energy as in the previous models. The model can reproduce the time scale of the first peak as well as the luminosity in the different bands (with the exception of the $i$ band at 5 days). The best fit implies $R_{ext}$~=~1200~$R_{\odot}$, $M_{ext}$~=~0.05~$M_{\odot}$ and an explosion epoch occurring 0.7 observer-frame days before the last-non detection, as this is deep enough only to catch an emission brighter than $-16.5$ mag in the $g$ band. 

There is a degeneracy between the two parameters $M_{ext}$ and $R_{ext}$, as shown in the bottom panel of Fig.~\ref{nakar}. From this plot, we can conclude that around the progenitor of iPTF15dtg there was an envelope with a radius of at least $\gtrsim$500~$R_{\odot}$ and with a mass of $\gtrsim0.045~M_{\odot}$. We note that for SN~IIb 2013cu (iPTF13ast, \citealp{galyam14}) the wind/envelope radius was estimated to be $\gtrsim$~720~$R_{\odot}$, similar to our object.

\section{Discussion}
\label{sec:discussion}

Based on the fits of the early light curve, we favored the extended envelope scenario over the interaction with a companion star. 
The interaction scenario is also disfavored by the first two spectra, which show no interaction signatures.
These spectra
were taken before $+$5 d, when the early decline would still be powered by this interaction. Relatively broad lines arising from the ejecta region where the reverse shock is propagating after the interaction with the companion should be visible in the high S/N spectrum taken with the Keck I telescope. We could also see narrow lines arising from the companion wind. However, the SN spectrum looks just like a normal SN~Ic spectrum and the observed narrow lines are from the host galaxy. 

We also considered the possibility that the early peak of iPTF15dtg was the result of an afterglow following a GRB or X-ray flash.
 However, the early spectrum does not resemble the afterglow spectra observed for GRB SNe. Furthermore, our radio upper limits constrain the 6 GHz radio flux of this SN to be $\leq$ 2$\times$10$^{27}$ erg~s$^{-1}$~Hz$^{-1}$. This is an order of magnitude fainter that the emission from the GRB-associated SN~1998bw \citep{kulkarni98} at comparable epochs (40$-$60~d since explosion). 

As mentioned in the introduction, SN~Ic-BL 2006aj was associated with an X-ray flash and showed an early declining phase which was interpreted as the signature of an extended envelope, as in the case of iTPF15dtg. Furthermore, the metallicity
 at the location of iPTF15dtg is compatible with those of long-duration GRBs and the ejecta mass and $^{56}$Ni mass of iPTF15dtg are in agreement with GRB-SNe \citep{cano16}. 
  However, iPTF15dtg also shows clear differences from SNe~Ic-BL associated to GRBs. More specifically, the main differences between iPTF15dtg and GRB-SNe are: (i) the kinetic energy is too 
low for a typical GRB-SN (even though \citealp{mazzali06} found $E_{K}$ $=$ $2\times$10$^{51}$ erg for SN~2006aj, 
similar to that we found here). (ii) No broad lines are observed in the optical spectra of iPTF15dtg and this certainly 
implies a normal Type Ic. However, we note that there was also a GRB-SN for which the features were not very
 broad (GRB 130215A /SN~2013ez; \citealp{cano14}), with derived line velocities comparable to those of normal SNe~Ic (4000--6000 km~s$^{-1}$ near peak light) rather than SNe~Ic-BL/GRB-SNe. 
  Despite a few similarities with SN~2006aj, PTF15dtg is not associated with a GRB/X-ray flash, and this could be
due to the fact that iPTF15dtg does not have or has a weaker central engine. This weak central engine might not be able to power a jet that breaks through the SN ejecta as in the case of GRB-SNe \citep{wang08}. However, a weaker central engine might be
enough to produce the high degree of $^{56}$Ni mixing that we observed in iPTF15dtg. 

Our early light curve models depend on the fit of the main peak that we performed with hydrodynamical models. One limitation is that we exploded a He-rich star to produce the bolometric light curve of iPTF15dtg, despite the fact that this SN is likely He poor. This is due to the difficulty of completely stripping the He layer with the current evolutionary models of single stars. However, we note that the He fraction of the progenitor stars of SE SNe should not significantly affect the bolometric light curve from the hydrodynamical model, as shown by \citet{dessart16} with their He-rich and He-poor models from binaries. The parameter of the progenitor star that mainly affects the SN bolometric light curve is its final mass, which we tuned to be $\sim$12 $M_{\odot}$. Another caveat concerning the hydrodynamical model is that SNEC is a 1D code and therefore we cannot take into account geometries that are different from spherical symmetry.  \citet{taubenberger09} and \citet{valenti12} indicate that a signature of bipolar explosions, typical of SNe with central engines, is a double peak in the profile of the nebular oxygen lines (even though the absence of a double peak does not rule out asphericity). We inspected the [\ion{O}{i}]~$\lambda\lambda$6300,6364 line in our last spectrum (which is not fully nebular) and the double peak is not visible. Another limitation of our hydrodynamical model is that it is not fit to the very early emission (cooling phase), but its parameters are used as input for the analytic models used to reproduce the early decline. 
Furthermore, to fit the bolometric light curve we do not explore the entire parameter space, but we restrict the range of the parameters that we investigate by using the Arnett model and the metallicity measurement.

The presence of material around the SN progenitor (deduced 
from the early emission), and the large ejecta mass (deduced from the slow rise to the main peak) both point to a progenitor star with large initial mass. Such a star (e.g., a WR star with $M_{ZAMS}\gtrsim35~M_{\odot}$ ) would be stripped of its outer layers by strong winds but would still leave $\sim$10 $M_{\odot}$ of ejecta, as estimated for iPTF15dtg. Furthermore, a massive WR star could produce an extended CSM envelope through episodic mass-loss events occurring prior to collapse. A massive eruptive episode from a WR star just a few years before collapse has been observed in the case of the progenitor of SN~2006jc \citep[][but see also \citealp{corsi14} on PTF11qcj]{pastorello07}.

If the 0.05$M_{\odot}$ of the $R~=~1200~R_{\odot}$ extended envelope of iPTF15dtg are
formed via an eruption, assuming a wind velocity of $\sim$1000~km~s$^{-1}$ (typical of massive WR stars) the envelope might have formed in a eruptive phase $\sim$10 days before collapse and characterized by extremely high mass-loss rate ($\dot{M}~=~1.9~M_{\odot}~$yr$^{-1}$). We stress that the extension and the mass of the envelope are lower limits (see Fig.~\ref{nakar}. If we assume an envelope radius two times that of the best fit with a mass of 0.055~$M_{\odot}$ (which is still a good fit, see the bottom panel of Fig.~\ref{nakar}), then the derived mass-loss rate would be about 1.0~$M_{\odot}~$yr$^{-1}$. 
 If a large mass-ejection occurred in the weeks before explosion, this could have produced a pre-explosion outburst observable in the light curve, as in the case of SN~2006jc. The eruption event of SN~2006jc was observed at $\sim$4 mag below the main peak \citep{pastorello07}, peaking at $\sim-14$~mag.
In the case of iPTF15dtg, our pre-explosion limits in $g$ band are not deep enough to detect such a faint outburst
and they only cover limited time intervals between $-$75~d and 0~d. Furthermore, the extended envelope mass of iPTF15dtg is markedly lower than the shell ejected by the precursor of SN~2006jc (which was a strongly interacting SN~Ibn) and hence the outburst of the progenitor for iPTF15dtg would have been even fainter.
 
The envelope properties of iPTF15dtg are more similar to those
estimated for the mass surrounding SN~IIb~2013cu (iPTF13ast) \citep{galyam14}, with a CSM extending at least $\sim$720 $R_{\odot}$ containing $\sim$0.004-0.017$ M_{\odot}$,  produced by a mass-loss rate of $>$ 0.03~$M_{\odot}$~yr$^{-1}$. Similarly, \citet{nakar15_06aj} found the progenitor of SN~2006aj to be surrounded by an envelope larger than 100~$R_{\odot}$ and containing 0.01~$M_{\odot}$.
In SN~2013cu the presence of the envelope was marked by strong  flash-spectroscopy emission lines which disappeared sometime after 3.23~d but before 6.45~d \citep{galyam14}. Our first spectra for iPTF15dtg were obtained at 3.2$\pm$0.5~d and 4.5$\pm$0.5~d, and they do not show these features. 

There could be another explanation for the extended envelope. If the progenitor of iPTF15dtg was part of a binary system, and the progenitor explodes during a common envelope phase \citep{yoon15}, this would imply a larger progenitor radius, consistent with the presence of a large envelope. The lack of flash spectroscopy signatures could also signal a different density profile of the SN environment, potentially due to binary evolution.

\begin{figure}
\centering
\includegraphics[width=9cm]{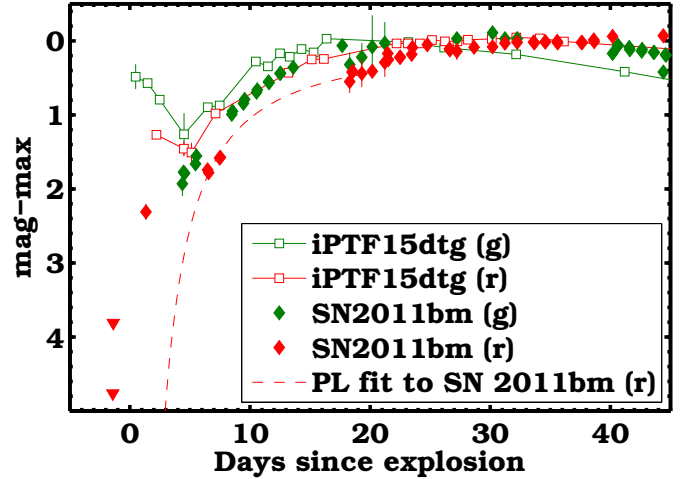}
\caption{\label{comp11bm}Early time light curves ($g$ and $r$ band) of iPTF15dtg (empty symbols) and SN~2011bm($=$PTF11bov, filled symbols), with data from PTF and \citet{valenti12}. Light curves are scaled to maximum. The early  $r$-band data points reveal that there could be an excess in the emission of SN~2011bm, similar to what is observed for iPTF15dtg. The best power-law fit to the early data of SN~2011bm (excluding the first epoch) is shown by a dashed line.
}
\end{figure}

If extended envelopes are characteristic of massive WR progenitors of SNe~Ic, we should find signatures of an early declining phase in the light curves of other long-rising SNe~Ibc, which presumably have similar progenitors. For instance, we saw that iPTF15dtg is quite similar to SN~2011bm, both photometrically and spectroscopically.
The light curves of SN~2011bm presented by \citet{valenti12} cover only one early epoch in the $r$ band, and thus it was difficult to asses the presence of an early declining phase. However, SN~2011bm was also observed by PTF (SN~2011bm$=$PTF11bov), with good coverage at early epochs in both $g$ and $r$ band. In Fig.~\ref{comp11bm} we combine the data from \citet{valenti12} and PTF in $r$ band (and also in $g$ band), and it appears that the first $r-$band point of SN~2011bm  displays an excess compared to the expected power-law (PL) rise (red dashed lines) obtained by fitting the early PTF $r-$band data points. Also in the case of SN~2011bm we thus have an early emission similar to that of iPTF15dtg, although with higher contrast with respect to the main peak. For the $g$ band it is not possible to assess the presence of an early declining phase with a time scale similar to that of iPTF15dtg as the first observations occurred too late.  For other (faster rising) SNe~Ic this early excess is not present, as in the cases of SN~1994I \citep{sauer06}, PTF10vgv (\citealp{corsi12}, \citealp{piro15}) or the SNe~Ic presented by \citet{taddia15}. The early excess could be a common property of spectroscopically normal SNe~Ic with slow rise, i.e., SN~2011bm-like events. 

Slow rising SNe~Ic (from massive WR stars) might be considered as bridging-gap objects between normal SNe~Ib/c and SLSNe~I, since their initial and ejecta masses are likely in-between those of the other two classes \citep{nicholl15_ejecta}. Normal SNe~Ib/c may need the help of a companion to get rid of their H/He rich envelopes, whereas slow-rising SNe~Ic and SLSN I might suffer large enough mass-loss due to their stronger winds. Slow-rising SNe~Ic such as iPTF15dtg and SN~2011bm show early peaks as do SLSNe~I \citep{nicholl15}, but the former are characterized by shorter time scales. As the time scale of the peaks are mainly dependent on the envelope mass, this implies larger envelope masses for SLSNe as compared to those of slow rising SNe~Ic.

\section{Conclusions}
\label{sec:conclusions}

\begin{itemize}
\item{iPTF15dtg is the first spectroscopically-normal SN~Ic showing an early declining phase in the optical light curves.}
\item{This SN also shows a long rise time, comparable only to that of SN~2011bm among SNe~Ic.}
\item{Analytic and hydrodynamical models of the main  peak as well as the fit to the expansion velocities reveal a large ejecta mass of $\sim$10~$M_{\odot}$, with an explosion energy of $\sim$2$\times$10$^{51}$~erg.}
\item{The modelling of the early light curve decline favours the presence of an extended envelope surrounding the progenitor star. Shock breakout from the SN progenitor is excluded, whereas companion interaction or magnetar models are not favored.}
\item{The large ejecta mass and the presence of extended material around the star suggest the progenitor star of iPTF15dtg was a massive Wolf-Rayet star, whose strong winds drove the stripping of the outer layers. It cannot be excluded that the WR star was in a binary system and exploded during a common envelope phase leading to the formation of the extended envelope.}
\item{SN~2011bm is found to show an early excess similar to that of iPTF15dtg. This might be a common characteristic of slow rising SNe~Ic.}
\end{itemize}

\begin{acknowledgements}
We gratefully acknowledge the support from the Knut and Alice Wallenberg Foundation. 

This work is partly based on observations made with the Nordic Optical Telescope, operated by the Nordic Optical Telescope Scientific Association at the Observatorio del Roque de los Muchachos, La Palma, Spain, of the Instituto de Astrofisica de Canarias.

The data presented here were obtained [in part] with ALFOSC, which is provided by the Instituto de Astrofisica de Andalucia (IAA) under a joint agreement with the University of Copenhagen and NOTSA.

This work is partly based on observations made with DOLoRes@TNG.

This paper made use of Lowell Observatoryʼs Discovery Channel  Telescope (DCT). Lowell operates the DCT in partnership with Boston University, Northern Arizona Uni- versity, the University of Maryland, and the University of Toledo. Partial support of the DCT was provided by Discovery Communications. The Large Monolithic Imager (LMI) on DCT was built by Lowell Observatory using funds from the National Science Foundation (AST-1005313).

LANL participation in iPTF was funded by the US Department of Energy as part of the Laboratory Directed Research and Development program.
 
Part of this research was carried out at the Jet Propulsion Laboratory, California Institute of Technology, under a contract with the National Aeronautics and Space Administration.

We thank N. Blagorodnova, E. Bellm, Y. Cao, G. Duggan, S. Kulkarni, J. Jencson, P. Nugent, for their precious help with the observations of iPTF15dtg and contribution to iPTF. We thank L. Yan for her comments on the paper.

Based  on  observations  obtained  with  the  Samuel  Oschin
Telescope 48-inch and the 60-inch Telescope at the Palomar
Observatory  as  part  of  the  intermediate  Palomar  Transient
Factory (iPTF) project,  a scientific collaboration among the
California Institute of Technology, Los Alamos National Lab-
oratory, the University of Wisconsin, Milwaukee, the Oskar
Klein Center, the Weizmann Institute of Science, the TANGO
Program of the University System of Taiwan, and the Kavli
Institute  for  the  Physics  and  Mathematics  of  the  Universe.

\end{acknowledgements}

\bibliographystyle{aa}

\begin{thebibliography}{100}

\expandafter\ifx\csname natexlab\endcsname\relax\def\natexlab#1{#1}\fi

\bibitem[Ahn et al.(2014)]{ahn14} Ahn, C.~P., Alexandroff, 
R., Allende Prieto, C., et al.\ 2014, \apjs, 211, 17 

\bibitem[Arcavi et al.(2011)]{arcavi11} Arcavi, I., Gal-Yam, A., 
Yaron, O., et al.\ 2011, \apjl, 742, L18 

\bibitem[Arcavi et al.(2010)]{arcavi10} Arcavi, I., Gal-Yam, A., 
Kasliwal, M.~M., et al.\ 2010, \apj, 721, 777 

\bibitem[Ben-Ami et al.(2014)]{benami14} Ben-Ami, S., Gal-Yam, A., Mazzali, P.~A., et al.\ 2014, \apj, 785, 37 

\bibitem[Bersten et al.(2013)]{bersten13} Bersten, M.~C., Tanaka, M., Tominaga, N., Benvenuto, O.~G., \& Nomoto, K.\ 2013, \apj, 767, 143 

\bibitem[Bianco et al.(2014)]{bianco14} Bianco, F.~B., Modjaz, M., Hicken, M., et al.\ 2014, \apjs, 213, 19 

\bibitem[Brown et al.(2009)]{brown09} Brown, P.~J., Holland, S.~T., Immler, S., et al.\ 2009, \aj, 137, 4517 

\bibitem[Cano et al.(2016)]{cano16} Cano, Z., Wang, S.-Q., Dai, Z.-G., \& Wu, X.-F.\ 2016, arXiv:1604.03549 

\bibitem[Cano et al.(2014)]{cano14} Cano, Z., de Ugarte Postigo, A., Pozanenko, A., et al.\ 2014, \aap, 568, A19 

\bibitem[Cano(2013)]{cano13} Cano, Z.\ 2013, \mnras, 434, 1098 

\bibitem[Cano et al.(2011)]{cano11} Cano, Z., Bersier, D., Guidorzi, C., et al.\ 2011, \apj, 740, 41 

\bibitem[Campana et al.(2006)]{campana06} Campana, S., Mangano, V., Blustin, A.~J., et al.\ 2006, \nat, 442, 1008 

\bibitem[Cenko et al.(2006)]{cenko06} 
Cenko, S.~B., Fox, D.~B., Moon, D.-S., et al.\ 2006, \pasp, 118, 1396 

\bibitem[Chevalier(1982)]{chevalier82} Chevalier, R.~A.\ 1982, \apj, 258, 790 



\bibitem[Clocchiatti et al.(2011)]{clocchiatti11} Clocchiatti, A., 
Suntzeff, N.~B., Covarrubias, R., \& Candia, P.\ 2011, \aj, 141, 163 

\bibitem[Corsi et al.(2014)]{corsi14} Corsi, A., Ofek, E.~O., Gal-Yam, A., et al.\ 2014, \apj, 782, 42 

\bibitem[Corsi et al.(2012)]{corsi12} Corsi, A., Ofek, E.~O., Gal-Yam, A., et al.\ 2012, \apjl, 747, L5 

\bibitem[Djupvik 
\& Andersen(2010)]{djupvik10} Djupvik, A.~A., \& Andersen, J.\ 2010, Astrophysics and Space Science Proceedings, 14, 211 

\bibitem[D'Elia et 
al.(2015)]{delia15} D'Elia, V., Pian, E., Melandri, A., et al.\ 2015, \aap, 577, A116 

\bibitem[Dessart et al.(2016)]{dessart16} Dessart, L., Hillier, 
D.~J., Woosley, S., et al.\ 2016, \mnras, 

\bibitem[Drout et al.(2015)]{drout15} Drout, M.~R., 
Milisavljevic, D., Parrent, J., et al.\ 2015, arXiv:1507.02694 

\bibitem[Drout et al.(2011)]{drout11} Drout, M.~R., Soderberg, 
A.~M., Gal-Yam, A., et al.\ 2011, \apj, 741, 97 

\bibitem[Eldridge et al.(2013)]{eldridge13} Eldridge, J.~J., Fraser, M., Smartt, S.~J., Maund, J.~R., \& Crockett, R.~M.\ 2013, \mnras, 436, 774 



\bibitem[Ergon et 
al.(2014)]{ergon14} Ergon, M., Sollerman, J., Fraser, M., et al.\ 2014, \aap, 562, A17 


\bibitem[Folatelli et al.(2006)]{folatelli06} Folatelli, G., Contreras, C., Phillips, M.~M., et al.\ 2006, \apj, 641, 1039 

\bibitem[Filippenko(1997)]{filippenko97} Filippenko, A.~V.\ 1997, \araa, 35, 309 

\bibitem[Filippenko et al.(1995)]{filippenko95} Filippenko, A.~V., 
Barth, A.~J., Matheson, T., et al.\ 1995, \apjl, 450, L11 

\bibitem[Fremling et al.(2016)]{fremling16}
Fremling et al. in 2016, submitted.

\bibitem[Fukugita et al.(1996)]{fukugita96} Fukugita, M., Ichikawa, T., Gunn, J.~E., et al.\ 1996, \aj, 111, 1748 

\bibitem[Gal-Yam et al.(2014)]{galyam14} Gal-Yam, A., Arcavi, I., Ofek, E.~O., et al.\ 2014, \nat, 509, 471 


\bibitem[Gal-Yam(2012)]{galyam12} Gal-Yam, A.\ 2012, Science, 337, 927 


\bibitem[Hunter et 
al.(2009)]{hunter09} Hunter, D.~J., Valenti, S., Kotak, R., et al.\ 2009, \aap, 508, 371 

\bibitem[Inserra et al.(2013)]{inserra13} Inserra, C., Smartt, 
S.~J., Jerkstrand, A., et al.\ 2013, \apj, 770, 128 

\bibitem[Irwin \& Chevalier(2015)]{irwin15} Irwin, C.~M., \& Chevalier, R.~A.\ 2015, arXiv:1511.00336 


\bibitem[Kasen et al.(2015)]{kasen15} Kasen, D., Metzger, 
B.~D., \& Bildsten, L.\ 2015, arXiv:1507.03645 



\bibitem[Kasen(2010)]{kasen10} Kasen, D.\ 2010, \apj, 708, 1025 


\bibitem[Kasen 
\& Bildsten(2010)]{kasen10_mag} Kasen, D., \& Bildsten, L.\ 2010, \apj, 717, 245 

\bibitem[Komatsu et al.(2009)]{komatsu09} Komatsu, E., Dunkley, 
J., Nolta, M.~R., et al.\ 2009, \apjs, 180, 330 

\bibitem[Kr{\"u}hler et al.(2015)]{kruhler15} Kr{\"u}hler, T., Malesani, D., Fynbo, J.~P.~U., et al.\ 2015, \aap, 581, A125 


\bibitem[Kulkarni et al.(1998)]{kulkarni98} Kulkarni, S.~R., Frail, D.~A., Wieringa, M.~H., et al.\ 1998, \nat, 395, 663 

\bibitem[Leloudas et al.(2012)]{leloudas12} Leloudas, G., Chatzopoulos, E., Dilday, B., et al.\ 2012, \aap, 541, A129 

\bibitem[Lunnan et al.(2014)]{lunnan14} Lunnan, R., Chornock, R., Berger, E., et al.\ 2014, \apj, 787, 138 

\bibitem[Lyman et al.(2016)]{lyman16} Lyman, J.~D., Bersier, 
D., James, P.~A., et al.\ 2016, \mnras, 457, 328 

\bibitem[Lyman et al.(2014)]{lyman14} Lyman, J.~D., Bersier, 
D., \& James, P.~A.\ 2014, \mnras, 437, 3848 

\bibitem[Maeda et al.(2007)]{maeda07} Maeda, K., Tanaka, M., Nomoto, K., et al.\ 2007, \apj, 666, 1069 


\bibitem[Malesani et al.(2009)]{malesani09} Malesani, D., Fynbo, 
J.~P.~U., Hjorth, J., et al.\ 2009, \apjl, 692, L84 

\bibitem[Margutti et al.(2015)]{margutti15} Margutti, R., Guidorzi, C., Lazzati, D., et al.\ 2015, \apj, 805, 159 

\bibitem[Mazzali et al.(2008)]{mazzali08} Mazzali, P.~A., Valenti, S., Della Valle, M., et al.\ 2008, Science, 321, 1185 

\bibitem[Mazzali et al.(2006)]{mazzali06} Mazzali, P.~A., Deng, J., Nomoto, K., et al.\ 2006, \nat, 442, 1018 


\bibitem[Modjaz et al.(2015)]{modjaz15} Modjaz, M., Liu, Y.~Q., 
Bianco, F.~B., \& Graur, O.\ 2015, arXiv:1509.07124 

\bibitem[Modjaz et al.(2014)]{modjaz14} Modjaz, M., Blondin, S., Kirshner, R.~P., et al.\ 2014, \aj, 147, 99
 
\bibitem[Modjaz et al.(2009)]{modjaz09} Modjaz, M., Li, W., 
Butler, N., et al.\ 2009, \apj, 702, 226 

\bibitem[Morozova et al.(2015)]{morozova15} Morozova, V., Piro, A.~L., Renzo, M., et al.\ 2015, \apj, 814, 63 

\bibitem[Nakar(2015)]{nakar15_06aj} Nakar, E.\ 2015, \apj, 807, 172 

\bibitem[Nakar 
\& Piro(2014)]{nakar14} Nakar, E., \& Piro, A.~L.\ 2014, \apj, 788, 193 

\bibitem[Nicholl 
\& Smartt(2016)]{nicholl16} Nicholl, M., \& Smartt, S.~J.\ 2016, \mnras, 457, L79 

\bibitem[Nicholl et al.(2015a)]{nicholl15} Nicholl, M., Smartt, 
S.~J., Jerkstrand, A., et al.\ 2015, \apjl, 807, L18 

\bibitem[Nicholl et al.(2015b)]{nicholl15_ejecta} Nicholl, M., Smartt, S.~J., Jerkstrand, A., et al.\ 2015, \mnras, 452, 3869 


\bibitem[Oke et al.(1995)]{oke95} Oke, J.~B., Cohen, J.~G., Carr, M., et al.\ 1995, \pasp, 107, 375 

\bibitem[Pastorello et al.(2007)]{pastorello07} Pastorello, A., Smartt, S.~J., Mattila, S., et al.\ 2007, \nat, 447, 829 

\bibitem[Paxton et al.(2011)]{paxton11} Paxton, B., Bildsten, 
L., Dotter, A., et al.\ 2011, \apjs, 192, 3 

\bibitem[Perley et al.(2011)]{perley09} Perley, R.~A., Chandler, C.~J., Butler, B.~J., \& Wrobel, J.~M.\ 2011, \apjl, 739, L1 

\bibitem[Pettini \& Pagel(2004)]{pp04} 
Pettini, M., \& Pagel, B.~E.~J.\ 2004, \mnras, 348, L59 

\bibitem[Piro(2015)]{piro15} Piro, A.~L.\ 2015, \apjl, 808, 
L51 

\bibitem[Piro 
\& Nakar(2013)]{piro13} Piro, A.~L., \& Nakar, E.\ 2013, \apj, 769, 67 


\bibitem[Prentice et al.(2016)]{prentice16} Prentice, S.~J., 
Mazzali, P.~A., Pian, E., et al.\ 2016, \mnras,  

\bibitem[Quimby et al.(2011)]{quimby11} Quimby, R.~M., Kulkarni, 
S.~R., Kasliwal, M.~M., et al.\ 2011, \nat, 474, 487 

\bibitem[Rahmer et al.(2008)]{rahmer08} 
Rahmer, G., Smith, R., Velur, V., et al.\ 2008, \procspie, 7014,  

\bibitem[Rabinak \& Waxman(2011)]{rabinak11} Rabinak, I., \& Waxman, E.\ 2011, \apj, 728, 63 


\bibitem[Richmond et al.(1996)]{richmond96} Richmond, M.~W., van 
Dyk, S.~D., Ho, W., et al.\ 1996, \aj, 111, 327 

\bibitem[Richmond et al.(1994)]{richmond94} Richmond, M.~W., Treffers, R.~R., Filippenko, A.~V., et al.\ 1994, \aj, 107, 1022 



\bibitem[Schlafly 
\& Finkbeiner(2011)]{sf11} Schlafly, E.~F., \& Finkbeiner, D.~P.\ 2011, \apj, 737, 103 

\bibitem[Sanders et al.(2013)]{sanders13} Sanders, N.~E., 
Levesque, E.~M., \& Soderberg, A.~M.\ 2013, \apj, 775, 125 



\bibitem[Sanders et al.(2012)]{sanders12} Sanders, N.~E., 
Soderberg, A.~M., Levesque, E.~M., et al.\ 2012, \apj, 758, 132 

\bibitem[Sauer et al.(2006)]{sauer06} Sauer, D.~N., Mazzali, P.~A., Deng, J., et al.\ 2006, \mnras, 369, 1939 

\bibitem[Smartt(2009)]{smartt09} Smartt, S.~J.\ 2009, \araa, 47, 63 

\bibitem[Soderberg et al.(2008)]{soderberg08} Soderberg, A.~M., 
Berger, E., Page, K.~L., et al.\ 2008, \nat, 453, 469 

\bibitem[Sollerman et al.(2006)]{sollerman06} Sollerman, J., Jaunsen, A.~O., Fynbo, J.~P.~U., et al.\ 2006, \aap, 454, 503 

\bibitem[Stritzinger et al.(2002)]{stritzinger02} Stritzinger, M., Hamuy, M., Suntzeff, N.~B., et al.\ 2002, \aj, 124, 2100 

\bibitem[Taddia et 
al.(2015)]{taddia15} Taddia, F., Sollerman, J., Leloudas, G., et al.\ 2015, \aap, 574, A60 

\bibitem[Taddia et 
al.(2015b)]{taddia15met} Taddia, F., Sollerman, J., Fremling, C., et al.\ 2015, \aap, 580, A131 

\bibitem[Taddia et al.(2013)]{taddia13met} 
Taddia, F., Sollerman, J., Razza, A., et al.\ 2013, \aap, 558, A143


\bibitem[Taubenberger et al.(2009)]{taubenberger09} Taubenberger, S., Valenti, S., Benetti, S., et al.\ 2009, \mnras, 397, 677

\bibitem[Taubenberger et al.(2006)]{taubenberger06} Taubenberger, S., 
Pastorello, A., Mazzali, P.~A., et al.\ 2006, \mnras, 371, 1459 


\bibitem[Valenti et al.(2012)]{valenti12} Valenti, S., 
Taubenberger, S., Pastorello, A., et al.\ 2012, \apjl, 749, L28 

\bibitem[Valenti et al.(2011)]{valenti11} Valenti, S., Fraser, 
M., Benetti, S., et al.\ 2011, \mnras, 416, 3138 


\bibitem[Valenti et al.(2008)]{valenti08} Valenti, S., 
Elias-Rosa, N., Taubenberger, S., et al.\ 2008, \apjl, 673, L155 


\bibitem[Wang \& Wheeler(2008)]{wang08} Wang, L., \& Wheeler, J.~C.\ 2008, \araa, 46, 433 

\bibitem[Waxman et al.(2007)]{waxman07} Waxman, E., M{\'e}sz{\'a}ros, P., \& Campana, S.\ 2007, \apj, 667, 351 


\bibitem[Wheeler et al.(2015)]{wheeler15} Wheeler, J.~C., Johnson, V., \& Clocchiatti, A.\ 2015, \mnras, 450, 1295 


\bibitem[Woosley \& Bloom(2006)]{woosley06} Woosley, S.~E., \& Bloom, J.~S.\ 2006, \araa, 44, 507 


\bibitem[Yaron \& Gal-Yam(2012)]{yaron12} Yaron, O., \& Gal-Yam, A.\ 2012, \pasp, 124, 668 



\bibitem[Yoon(2015)]{yoon15} Yoon, S.-C.\ 2015, \pasa, 32, e015 








\end{thebibliography}

\onecolumn

\onecolumn

\begin{deluxetable}{|cc|cc|cc|cc|}
\tabletypesize{\scriptsize}
\tablewidth{0pt}
\tablecaption{Optical photometry of iPTF15dtg.\label{tab:phot}}
\tablehead{
\colhead{JD-2,457,000}&
\colhead{$B$}&
\colhead{JD-2,457,000}&
\colhead{$g$}&
\colhead{JD-2,457,000}&
\colhead{$r$}&
\colhead{JD-2,457,000}&
\colhead{$i$}\\
\colhead{(days)}&
\colhead{(mag)}&
\colhead{(days)}&
\colhead{(mag)}&
\colhead{(days)}&
\colhead{(mag)}&
\colhead{(days)}&
\colhead{(mag)}}
\startdata
337.945   &  20.421(0.087)   &  333.931 & 19.634(0.162) &  335.660 & 19.968(0.026) &   335.662 & 20.308(0.037) \\ 
340.573   &  20.507(0.077)   &  334.931 & 19.720(0.029) &  337.950 & 20.156(0.097) &   337.953 & 20.671(0.122) \\ 
346.653   &  19.799(0.057)   &  335.931 & 19.943(0.027) &  338.586 & 20.208(0.129) &   340.589 & 20.499(0.100) \\ 
349.570   &  19.724(0.055)   &  337.948 & 20.383(0.082) &  340.586 & 19.680(0.061) &   346.664 & 19.465(0.036) \\ 
355.655   &  19.672(0.069)   &  337.960 & 20.412(0.278) &  346.661 & 19.130(0.002) &   348.582 & 19.324(0.010) \\ 
356.591   &  19.616(0.061)   &  338.584 & 20.288(0.109) &  348.580 & 18.950(0.005) &   349.602 & 19.257(0.028) \\ 
357.589   &  19.736(0.049)   &  339.933 & 20.046(0.012) &  349.599 & 18.944(0.038) &   355.655 & 19.084(0.008) \\ 
358.589   &  19.633(0.094)   &  340.584 & 20.028(0.071) &  355.652 & 18.734(0.019) &   356.598 & 19.027(0.027) \\ 
359.669   &  19.746(0.030)   &  340.936 & 20.022(0.015) &  356.595 & 18.735(0.029) &   357.599 & 19.018(0.024) \\ 
360.712   &  19.757(0.001)   &  343.941 & 19.427(0.012) &  357.597 & 18.735(0.011) &   358.596 & 18.977(0.033) \\ 
          &                  &  344.941 & 19.494(0.049) &  358.594 & 18.688(0.046) &   359.687 & 18.946(0.020) \\ 
          &                  &  345.951 & 19.318(0.021) &  359.684 & 18.706(0.007) &   360.721 & 18.917(0.016) \\ 
          &                  &  346.659 & 19.392(0.027) &  360.718 & 18.706(0.020) &   361.592 & 18.911(0.052) \\ 
          &                  &  346.759 & 19.363(0.005) &  361.590 & 18.685(0.036) &   363.582 & 18.908(0.027) \\ 
          &                  &  347.738 & 19.259(0.004) &  363.580 & 18.675(0.031) &   365.616 & 18.857(0.026) \\ 
          &                  &  348.578 & 19.293(0.013) &  365.613 & 18.654(0.019) &   367.584 & 18.846(0.053) \\ 
          &                  &  348.750 & 19.306(0.050) &  367.582 & 18.666(0.051) &   369.581 & 18.896(0.030) \\ 
          &                  &  349.597 & 19.268(0.045) &  369.578 & 18.707(0.034) &   380.421 & 19.051(0.035) \\ 
          &                  &  349.827 & 19.120(0.053) &  380.419 & 18.831(0.038) &   386.443 & 19.031(0.025) \\ 
          &                  &  355.650 & 19.207(0.015) &  386.442 & 18.916(0.014) &   393.355 & 19.087(0.026) \\ 
          &                  &  356.593 & 19.187(0.036) &  393.351 & 19.041(0.024) &   398.357 & 19.274(0.041) \\ 
          &                  &  356.634 & 19.165(0.001) &  398.353 & 19.142(0.042) &   400.424 & 19.244(0.043) \\ 
          &                  &  357.595 & 19.242(0.001) &  400.421 & 19.186(0.014) &   410.417 & 19.213(0.059) \\ 
          &                  &  358.591 & 19.245(0.055) &  410.413 & 19.415(0.054) &   413.372 & 19.440(0.023) \\ 
          &                  &  359.615 & 19.240(0.010) &  413.366 & 19.344(0.019) &   413.623 & 19.429(0.001) \\ 
          &                  &  359.682 & 19.284(0.007) &  413.618 & 19.292(0.001) &   416.639 & 19.472(0.001) \\ 
          &                  &  360.716 & 19.307(0.008) &  416.634 & 19.329(0.001) &   420.678 & 19.442(0.001) \\ 
          &                  &  361.588 & 19.257(0.046) &  420.663 & 19.328(0.004) &   423.679 & 19.562(0.007) \\ 
          &                  &  363.578 & 19.331(0.037) &  423.674 & 19.377(0.001) &   441.619 & 19.784(0.003) \\
          &                  &  365.604 & 19.395(0.030) &  441.614 & 19.630(0.002) &   444.701 & 19.801(0.001) \\
          &                  &  365.612 & 19.332(0.010) &  444.696 & 19.724(0.001) &   448.639 & 19.877(0.001) \\
          &                  &  367.579 & 19.391(0.069) &  448.625 & 19.697(0.002) &   451.638 & 19.960(0.084) \\
          &                  &  369.576 & 19.475(0.050) &  451.635 & 19.650(0.070) &   458.355 & 19.905(0.022) \\
          &                  &  374.636 & 19.565(0.021) &  458.349 & 19.713(0.024) &   463.378 & 19.838(0.031) \\
          &                  &  380.417 & 19.942(0.088) &  463.372 & 19.697(0.026) &           &               \\
          &                  &  386.439 & 19.944(0.022) &          &               &           &               \\
          &                  &  389.634 & 19.998(0.030) &          &               &           &               \\
          &                  &  393.347 & 20.099(0.024) &          &               &           &               \\
          &                  &  398.348 & 20.177(0.045) &          &               &           &               \\
          &                  &  400.416 & 20.292(0.022) &          &               &           &               \\
          &                  &  401.625 & 20.195(0.032) &          &               &           &               \\
          &                  &  410.407 & 20.414(0.093) &          &               &           &               \\
          &                  &  413.359 & 20.526(0.043) &          &               &           &               \\
          &                  &  413.615 & 20.411(0.002) &          &               &           &               \\
          &                  &  416.629 & 20.422(0.003) &          &               &           &               \\
          &                  &  423.669 & 20.465(0.006) &          &               &           &               \\
          &                  &  424.635 & 20.424(0.022) &          &               &           &               \\
          &                  &  427.655 & 20.434(0.243) &          &               &           &               \\
          &                  &  430.639 & 20.376(0.089) &          &               &           &               \\
		  &                  &  444.691 & 20.645(0.001) &          &               &           &               \\
		  &                  &  448.620 & 20.683(0.001) &          &               &           &               \\
		  &                  &  451.642 & 20.620(0.068) &          &               &           &               \\
		  &                  &  458.340 & 20.727(0.032) &          &               &           &               \\
		  &                  &  463.356 & 20.768(0.032) &          &               &           &               \\
  \enddata
\end{deluxetable}

\begin{deluxetable}{cccccc}
\tabletypesize{\scriptsize}
\tablewidth{0pt}
\tablecaption{Optical Spectroscopy of iPTF15dtg\label{tab:spectra}}
\tablehead{
\colhead{Date (UT)}&
\colhead{JD-2,457,000}&
\colhead{Phase\tablenotemark{a}}&
\colhead{Telescope}&
\colhead{Instrument}&
\colhead{Range}\\
\colhead{}&
\colhead{(days)}&
\colhead{(days)}&
\colhead{}&
\colhead{}&
\colhead{(\AA)}}
\startdata

 10 Nov 2015  & 336.60  &  +3.2 & TNG   &   DOLORES    & 3435$-$8076 \\
 11 Nov 2015  & 337.98  &  +4.5 & Keck1® &   LRIS       & 3074$-$10278 \\
 23 Nov 2015  & 350.46  & +17.0 & TNG   &   DOLORES    & 3561$-$10462 \\
 02 Dec 2015  & 358.85  & +25.4 & DCT   &   Deveny+LMI & 3272$-$7731 \\
 06 Dec 2015  & 362.50  & +29.1 & Keck1 &   LRIS       & 3070$-$10232 \\
 10 Dec 2015  & 367.43  & +34.0 & NOT   &   ALFOSC     & 3482$-$9083 \\
 10 Jan 2016  & 398.32  & +64.9 & TNG   &   DOLORES    & 3382$-$9600 \\
 28 Jan 2016  & 416.37  & +82.9 & NOT   &   ALFOSC     & 3463$-$9143 \\
 11 Feb 2016  & 430.42  & +97.0 & TNG   &   DOLORES    & 3371$-$9609 \\
 09 Mar 2016  & 456.73  & +123.3 & Gemini North  &   GMOS    & 3796$-$9346 \\
\enddata
\tablenotetext{a}{From explosion date.}
\end{deluxetable}

\end{document}